\documentclass[preprint,3p]{elsarticle}

\usepackage{amssymb}
\biboptions{sort&compress}
\usepackage{amsmath}
\usepackage[colorlinks,linkcolor=blue,anchorcolor=blue,citecolor=blue]{hyperref}
\usepackage{booktabs, array, mathptmx, float, tabularx, booktabs, lipsum, multirow,tabularx}
\usepackage{subfigure}
\usepackage{caption}
\usepackage{bm}

\usepackage[font=small,labelfont=bf,labelsep=none]{caption}  

\usepackage{cuted}
\UseRawInputEncoding

\journal{Nuclear Inst. and Methods in Physics Research, A}

\begin{document}
	
	\begin{frontmatter}
		
		
		
		\title{Simulation studies on compensation for incoherent magnet error driven half-integer and 3rd-order resonances with space charge in HIAF-BRing} 
		
		
		\author[1,2]{C.Guo}
		\author[1,2]{J.Liu}
		\author[1,2]{J.C.Yang\corref{cor1}}
		\cortext[cor1]{Corresponding author at: Institute of Modern Physics, Chinese Academy of Sciences, Lanzhou 730000, China.}
		\ead{yangjch@impcas.ac.cn}
		\author[1,2]{R.H.Zhu}
		
		\affiliation[1]{
			organization={Institute of Modern Physics, Chinese Academy of Sciences},
			city={Lanzhou},
			postcode={730000},
			country={China}
		}
		\affiliation[2]{
			organization={University of Chinese Academy of Sciences},
			city={Beijing},
			postcode={100049},
			country={China}
		}
		\begin{abstract}
			Magnet error driven resonances under space-charge-induced resonance crossing represents a significant limitation on beam intensity in low-injection-energy high-intensity synchrotrons. Unstable particle motion arises from the combined effects of space-charge-induced tune spread and magnetic field imperfections, which limits the intensity. This paper introduces a space-charge-Twiss modification to the Resonance Driving Terms (RDTs). The new RDTs are named modified RDTs. Modified RDTs aim to describe the nonlinear behavior under incoherent error driven resonances induced by the combined effect of space charge and magnetic field imperfections. The feasibility of the compensation scheme with modified RDTs is demonstrated through coasting-beam simulations under space-charge-induced half-integer and 3rd-order resonances, using the lattice of the High Intensity Heavy-Ion Accelerator Facility Booster Ring (HIAF-BRing). The simulations demonstrate that the compensation through minimizing modified RDTs significantly suppresses the emittance growth and the unstable particle motion. Besides, the effect of compensation scheme with modified RDTs against periodical resonance crossing is evaluated in bunched-beam simulations.
		\end{abstract}
		
		
		
		
	\end{frontmatter}
	
	
	
	\section{Introduction}
	\label{section1}
	
	Space charge effect is the key limitation of beam intensity in low-injection-energy high-intensity synchrotrons. It can lead to undesirable beam emittance growth, beam halo formation and particle loss. Since particles with different betatron amplitudes have different incoherent tune shifts, $\Delta\nu_{sc}(J_{x,y})$, the space charge force produces an incoherent tune spread. As the beam intensity gets stronger, the tune spread becomes larger, which would eventually lead to resonance crossing. Besides, the variation of the tunes through synchrotron motion due to the longitudinal dependency of space charge and chromatic effects leads to periodic resonance crossing. The space-charge-induced periodic resonance crossing has been identified as the main beam degradation mechanism for long storage times in low energy high-intensity synchrotrons, like Proton Synchrotron (PS) and the Super Proton Synchrotron (SPS) at CERN \cite{RN56}.
	
	Here are 4 approaches that have been proposed to mitigate this issue. Adrian Oeftiger et al. proposed a novel space charge compensation concept, pulsed electron lenses. This technique aims to mitigate space charge of ions through space charge of the electrons. Simulations of the Facility for Antiproton and Ion Research (FAIR) SIS100 synchrotron demonstrated that this method could achieve up to a 30\% increase in maximum beam intensity \cite{RN105}. 
	Hideaki Hotchi et al. at Facilities at Japan Proton Accelerator Research Complex (J-PARC) proposed a novel method of transverse injection painting to counteract the emittance exchange caused by Montague resonance. The beam stability was achieved by sextupole compensation under resonance crossing of 3rd-order resonance lines \cite{RN86}.  
	Meanwhile, the compensation scheme in the Main Ring (J-PARC-MR) was demonstrated in \cite{RN107}. 
	Andrea Santamaria Garcia and Foteini Asvesta et al. achieved the beam stability under resonance crossing of a half-integer resonance line by scanning the currents of quadrupole correctors at Proton Synchrotron Booster \cite{RN76,RN75}. 
	
	Resonance Driving Terms (RDTs \cite{RN35}) play an important role in resonance compensation. RDTs focus on identifying and quantifying the effects of specific perturbations that could induce resonant behavior of the beam. These perturbations include imperfections in the magnetic fields of the accelerator. The Hamiltonian of the system could be expressed as a series of Fourier coefficients depending on the amplitude and phase of the beam oscillations \cite{RN36}. In addition, RDTs from space charge potential were carried out in \cite{Asvesta2696190}. However, RDTs do not consider the influence of space charge on Twiss parameters, resulting in limited effectiveness in compensating for magnetic field imperfections under space-charge-induced incoherent error driven resonances. According to the analysis of Baartman \cite{RN52}, the core of the beam would not be affected by incoherent resonances. But according to the analysis of Hofmann \cite{RN121}, core particles would still be affected by incoherent resonances when Landau damping is taken into account. Incoherent error driven resonances play an important role in space charge accelerator physics. The main target of this paper is to establish an approach to suppress incoherent error driven resonances with space charge.
	
	The key idea of this paper is to propose the space-charge-Twiss modification in RDTs, aiming to mitigate the beam response to incoherent error driven resonances. The proposed modified RDTs are evaluated by coasting-beam simulations performed on a dedicated accelerator collective instability platform, CISP \cite{RN108}, using the lattice of the High Intensity heavy-ion Accelerator Facility Booster Ring (HIAF-BRing \cite{RN106}). Furthermore, bunched-beam simulations with synchrotron motion and chromaticity have also been performed. The compensation scheme of modified RDTs still had a good performance in the bunched-beam simulations. In these simulations, non-structural half-integer and 3rd-order resonance lines were crossed due to the space charge incoherent tune spread. Magnetic field errors were introduced to excite corresponding resonances. Modified RDTs were then used to determine the strengths of magnetic correctors in order to compensate for these errors. For comparison, compensation scheme through regular RDTs without modification is also demonstrated.
	
	The paper is structured as follows. In Sec.\ref{section2}, the space-charge-Twiss modification in RDTs is demonstrated. 
	In Sec.\ref{section3}, modified RDTs scheme and regular RDTs scheme are used in the simulation to compensate for corresponding magnetic field errors during resonance crossing. The half-integer and 3rd-order incoherent error driven resonances are successfully compensated by modified RDTs. The compensation still has a good performance while synchrotron motion is included in the simulations.
	Sec.\ref{section4} contains a summary of the whole paper.
	
	\section{\label{section2}Space-charge-Twiss modification in RDTs}
	
	RDTs, based on the description in normal forms of the beam oscillations, are quiet effective tools on nonlinear beam dynamics as shown in Eq.~(\ref{eq:x-ipx}). According to RDTs, the horizontal TbT data of the complex Courant-Snyder variables at the location of the Beam Position Monitor(BPM), denoted as $b$, can be decomposed into a series of Fourier components \cite{RN66}:
	\begin{equation}
		\begin{aligned}
			[\widehat{x} -i\widehat{p}_{x} ](N,b)\approx& 
			\sqrt{2I_{x}}e^{i(2\pi \nu _{x}N+ \phi _{x_{b} } )}  
			\\&
			-2i\sum_{jklm}^{}jf_{jklm}^{(b)}(2I_{x}) ^{\frac{j+k-1}{2} } (2I_{y} ) ^{\frac{l+m}{2} }
			\\&
			\times e^{i[(1-j+k)(2\pi \nu _{x}N+ \phi _{x_{b} })+(m-l)(2\pi \nu _{y}N+ \phi _{y_{b} })]}
			\label {eq:x-ipx}
		\end{aligned}
	\end{equation}
	where $\widehat{x}$ and $\widehat{p}_{x}$ are the normalized coordinates, $I_{x,y}$ and $\phi_{x,y}$ are the nonlinear action-angle variables, $I_{x,y}$ is the new invariant in the normal form space, $\phi_{x,y}$ is the new phase in the normal form space, and $\nu_{x,y}$ are the tune of the particle including the amplitude-dependent detuning\cite{RN63}. The horizontal and vertical phase of the lattice at the BPM location $b$ are denoted as $\phi _{x_{b},y_{b} }$.
	The resonance driving terms in Eq.~(\ref{eq:x-ipx}), $f_{jklm}^{(b)}$, are
	\begin{equation}
		f_{jklm}^{(b)}= 
		\frac{\sum_{\omega }^{}h_{\omega ,jklm}
			e^{i[(j-k)\triangle \phi _{\omega ,x}^{b}
				+(l-m)\triangle \phi _{\omega ,y}^{b} ]}}
		{1-e^{2\pi i[(j-k)Q_{x}+(l-m)Q_{y} ]} } 
		\label{eq:fjklm}
	\end{equation}
	where the $\omega$ is the location of the $\omega^{th}$ multipole, the $\triangle \phi _{\omega ,x,y}^{b}$ is the phase advance between the $\omega^{th}$ multipole and the location $b$, and the $h_{\omega ,jklm}$ is
	\begin{equation}
		\begin{aligned}
			&h_{\omega ,jklm}  = -\frac{\Omega _{\omega ,n-1}(l+m)}{j!k!l!m!2^{j+k+l+m} }i^{l+m}(\beta _{\omega ,x})^{\frac{j+k}{2}}(\beta _{\omega ,y})^{\frac{l+m}{2}} 
			\label{eq:hwjklm}
			\\
			&\Omega _{\omega ,n-1}(l+m)  = 
			\begin{cases}
				K_{\omega ,n-1}L &\text{if (l+m) is even} \\
				iKS_{\omega ,n-1}L &\text{if (l+m) is odd}
			\end{cases}
		\end{aligned}
	\end{equation}
	where $n$ is the order of resonance, $n=j+k+l+m$, $K$ is the normal multipole coefficient, $KS$ is the skew multipole coefficient and $L$ is the length of the $\omega^{th}$ multipole. When space charge effect could be ignored, the RDTs are useful tools to study the effect of magnetic nonlinear fields in the betatron motion. RDTs have a beautiful performance in many previous works related with transverse nonlinearity, such as \cite{RN117,RN87,RN85,RN66}. 
	
	Although RDTs from space charge potential have been investigated in \cite{Asvesta2696190}, previous works have not consider the influence of space-charge-induced single particle motion , especially the phase advance, on RDTs. As a result, RDTs are less effective in compensating for space-charge-induced error driven resonances, which is demonstrated in simulations of the next section. In the following text, space charge would be incorporated in order to study the combined effect of space charge and magnetic field imperfections, which leads to the incoherent error driven resonance.
	
	First of all it is necessary to understand the particle behavior with space charge and without magnet multipoles. When space charge could not be ignored, treating the space-charge potential $U_{sc}$ as a perturbation, the Hamiltonian becomes \cite{RN83}:
	\begin{equation}
		H=\frac{p_x^2+K_xx^2}{2} +\frac{p_y^2+K_yy^2}{2} +U_{sc}
		\label{eq_hamiltonlian}
	\end{equation}
	where the $K_{x,y}$ is the normalized quadrupole strength. It can be observed from Eq.~(\ref{eq_hamiltonlian}) that the total phase advance has 2 parts. The first part comes from the linear lattice, and the second part is the phase advance shift induced by space charge. This is recorded as:
	\begin{equation}
		\frac{d\phi_{sc}}{ds} = \frac{d\phi_{linear}}{ds} + \triangle\phi_{sc}
		\label{eq_two_parts_phase_advance}
	\end{equation}
	where $\phi_{sc}$ is the total phase advance with space charge, $\phi_{linear}$ is the phase advance induced by the linear lattice, and $\triangle\phi_{sc}$ is defined as the phase advance shift induced by space charge. In order to understand $\phi_{sc}$, $U_{sc}$ must be investigated. $U_{sc}$ is the space charge potential of a beam with arbitrary charge distribution with symmetry. For simplicity of analysis, one dimension is considered. $U_{sc}$ can be written as a summation of an infinite power series:
	\begin{equation}
		U_{sc}=\sum_{n=0}^{\infty } a_n(\frac{x}{\sigma_x})^{2n}
		\label{eq_arbitrary_space_charge_potential}
	\end{equation}
	where $a_n$ are coefficients which are dependent of the arbitrary beam charge distribution with symmetry, $\sigma_x$ is the RMS envelope radii. Detail dynamics about space charge dispersion can be found in \cite{dispersion}, but the contribution of dispersion on $\sigma_x$ is neglected for simplification. Substitute $\sigma_x = \sqrt{\varepsilon_{rms,x}\beta_x}$ into Eq.~(\ref{eq_arbitrary_space_charge_potential}):
	\begin{equation}
		U_{sc}=\sum_{n=0}^{\infty } a_n(\frac{1}{\varepsilon_{rms,x}})^n(\frac{x}{\sqrt{\beta_x}})^{2n}
		=\sum_{n=0}^{\infty}a_n(\frac{2}{\varepsilon_{rms,x}})^nJ_x^n\cos^{2n}{\phi_x }
		\label{eq_arbitrary_space_charge_potential_2}
	\end{equation}
	where the $\varepsilon_{rms,x}$ is the RMS emittance in x plane, $J_u=\frac{1}{2} (\gamma_uu^2+2\alpha _uup_u+\beta_u p_u^2)$ represents the action of particles and $\phi_x$ represents the angle of particles, $\alpha_u, \beta_u, \gamma_u$ are Twiss parameters. $x=\sqrt[]{2J_x\beta_x}\cos{\phi_x }$ is the action-angle variables. From the perspective of turn-by-turn oscillations, the angle variation pattern of each particle is determined by its tune. By further considering the cyclicity of the synchrotron, all particles would uniformly and repeatedly experience angles from 0 to 2$\pi$ at every longitudinal position. As a result, although the angle exists as a variable in Eq.~(\ref{eq_arbitrary_space_charge_potential_2}), it is reasonable to simplify the equation by integrating Eq.~(\ref{eq_arbitrary_space_charge_potential_2}) over the angle from 0 to 2$\pi$ to obtain its average. The same approach has also been applied in Ng's book\cite{RN65}:
	\begin{equation}
		\left \langle U_{sc} \right \rangle =\frac{1}{2\pi}\int_{0}^{2\pi}\sum_{n=0}^{\infty}a_n(\frac{2}{\varepsilon_{rms,x}})^nJ_x^n\cos^{2n}{\phi_x }d\phi_x  =\sum_{n=0}^{\infty}a_n(\frac{2}{\varepsilon_{rms,x}})^nJ_x^n\frac{1}{2\pi}\int_{0}^{2\pi} \cos^{2n}{\phi_x }d\phi_x
		\label{eq_average_Usc}
	\end{equation}
	where $\left \langle U_{sc} \right \rangle$ represents the average of $U_{sc}$ over angles. By extending the Wallis integral to the interval from [0, $\pi/2$] into [0, 2$\pi$], it can be obtained that:
	\begin{equation}
		\begin{aligned}
			\frac{1}{2\pi}\int_{0}^{2\pi}\cos^n{\phi_x } d\phi_x =\begin{cases}
				0 &\text{if n is odd}
				\\
				1  &\text{if n = 0}
				\\
				\frac{(n-1)!!}{n!!}  &\text{if n is even}
			\end{cases}
		\end{aligned}
		\label{eq_Wallis_integration}
	\end{equation}
	Substitute Eq.~(\ref{eq_Wallis_integration}) into Eq.~(\ref{eq_average_Usc}):
	\begin{equation}
		\left \langle U_{sc} \right \rangle 
		=a_0+\sum_{n=1}^{\infty}a_n(\frac{2}{\varepsilon_{rms,x}})^n \frac{(2n-1)!!}{(2n)!!}J_x^n
		\label{eq_average_Usc_2}
	\end{equation}
	With the action-angle variables in the averaged space charge potential $\left \langle U_{sc} \right \rangle $, the averaged phase advance shift induced by space charge, $\triangle \phi_{sc,x}$, can be obtained by:
	\begin{equation}
		 \triangle \phi_{sc,x} = \frac{\partial \left \langle U_{sc} \right \rangle}{\partial J_x} = \sum_{n=1}^{\infty}n(\frac{2}{\varepsilon_{rms,x}})^n\frac{(2n-1)!!}{(2n)!!}a_{n}J_x^{n-1}=\sum_{n=1}^{\infty}A_nJ_x^{n-1}
		 \label{eq_phase_advance}
	\end{equation}
	where $A_n=n(\frac{2}{\varepsilon_{rms,x}})^n\frac{(2n-1)!!}{(2n)!!}a_{n}$, which can just be treated as another group of coefficients which are determined by the arbitrary beam distribution. A regular approach to calculate $\triangle \phi_{sc,x}$ is firstly to obtain $J_x$ from the target particle, secondly to obtain $A_n$ from the beam distribution along the ring, and finally calculate $\triangle \phi_{sc,x}$ through Eq.~(\ref{eq_phase_advance}). This approach is precise but too complicated. A more convenient approach is established in the following text.
	
	The total tune shift induced by space charge, $\triangle\nu _{sc,x}$, can be obtained by integrating Eq.~(\ref{eq_phase_advance}) over the longitudinal position $s$ from 0 to C, which is the circumference of the synchrotron:
	\begin{equation}
		\triangle\nu _{sc,x} = \int_{0}^{C}\triangle\phi_{sc,x}ds 
		=\int_{0}^{C}\sum_{n=1}^{\infty}A_nJ_x^{n-1}ds
		\label{eq_tune_shift}
	\end{equation}
	The question is wether $A_n$ and $J_u$ is dependent or independent of $s$. $J_u$ is still invariant and independent of $s$ for core particles at the center of the beam because the space charge effect is almost linear in core region. After core particles transfer of an arbitrary longitudinal length, only their angles changes, which is a rotation in the normalized phase space. This rotation is amplitude-dependent due to the amplitude-dependent phase advance. Considering the symmetric distribution, particles with the same $J_u$ would uniformly distributed in each angle from 0 to $2\pi$. This means that amplitude-dependent rotation in would not change central beam charge distribution. As a result, $A_n$ is also independent of $s$ for core particles. However, $J_u$ is no longer invariant for halo particles at the edge of the beam, because the nonlinear space charge effect is serious in halo region. Another important issue in halo region, which can be approximately treated as a region outside the beam charge, is the contribution of $u^{-1}$ term on space charge potential, which mismatches the description of Eq.~(\ref{eq_arbitrary_space_charge_potential}). Unfortunately the $u^{-1}$ term can not be included in our analysis, because the integration Eq.~(\ref{eq_Wallis_integration}) is not convergent when $n=-1$. Besides, the space charge potential that halo particles experience is a piecewise function, where the $u^{-1}$ term is the main contribution when angles are located in halo region and Eq.~(\ref{eq_arbitrary_space_charge_potential}) is the main contribution in other cases. The analysis of halo particles with piecewise-function space charge potential awaits further study. As a result, it can be concluded that $A_n$ and $J_u$ is independent of $s$ for small-amplitude particles and the description of Eq.~(\ref{eq_arbitrary_space_charge_potential}) will be ineffective for halo particles. For non-halo particles, a key relationship of this paper can be obtained:
	\begin{equation}
		\frac{\triangle \phi_{sc,u}}{\triangle\nu _{sc,u}} = \frac{1}{C}
		\label{eq_tune_shift_phase_advance}
	\end{equation}
	where $u$ represents $x$ or $y$. The derivation in y plane is the same. According to the following simulation section, Eq.~(\ref{eq_tune_shift_phase_advance}) is effective except in the halo region. As a result, the total phase advance of particles modulated by space charge, $\phi_{sc,u}$, could be obtained through Eq.~(\ref{eq_tune_shift_phase_advance}):
	\begin{equation}
		\phi_{sc,u}(s) = \phi_{linear,u}(s)+\frac{s}{C}\triangle \nu_{sc,u} 
		\label{eq_sc_phase_advance}
	\end{equation}
	The phase advance of target particles under resonance crossing can be easily found through adjusting the value of $\Delta\nu_{sc,u}$. By choosing $\Delta\nu_{sc,u}$, the complicated approach through $A_n$ and $J_u$ can be substituted. 
	
	The first term of Eq.~(\ref{eq_arbitrary_space_charge_potential}), which represents the linear space charge strength, can be used in the RMS envelope equation\cite{4326293} to get $\beta_{sc,u}$ with space charge. The difference between the $\beta_{sc,u}$ and the linear $\beta_{u}$ is usually less than 1\%, which contributes little on the RDTs. As a result, the total phase advance of particles modulated by space charge, $\phi_{sc,u}$, is the key modification which should be applied in RDTs.
	
	The space charge treatment in this paper is to contain the space charge detuning. Although nonlinear space charge terms is included in space charge detuning, their nonlinear effect and corresponding resonances which might be driven are not considered. This simplification might raise issues on resonances that are driven at high order by space charge itself. Under the assumption that the nonlinear space charge effect is negligible, the differences between the space-charge-induced incoherent error driven resonances and the non-space-charge ones are the $\beta$ and $\phi$. According to the definition of RDTs formalism, the $\beta_{sc,u}$ and $\phi_{sc,u}$ with space charge should be considered in RDTs. As a result, the space-charge-induced incoherent error driven resonances behaviors can be obtained by replacing the $\beta_u$ and $\phi_{u}$ in RDTs with space charge induced $\beta_{sc,u}$ and $\phi_{sc,u}$:
	\begin{equation}
		f_{jklm}^{(b)}(\beta_u,\phi_{u})\longrightarrow f_{jklm}^{(b)}(\beta_{sc,u},\phi_{sc,u})
		\label{eq_modifiedRDTs}
	\end{equation}
	from which the space charge modification has been introduced into RDTs. Note that the $\beta_u$ in Eq.~(\ref{eq:hwjklm}) should be replaced as well. The beta function modulated by space charge, $\beta_{sc,u}$, has little influence on RDTs since the difference between $\beta_{sc,u}$ and $\beta_{u}$ is usually less than 1\% as previously discussed. But the phase modulated by space charge, $\phi_{sc,u}$, would make significant influence on RDTs, which is the main difference between regular RDTs and new RDTs. The new RDTs are named modified RDTs, which could be used to compensate for magnetic field imperfections under space-charge-induced incoherent error driven resonances. This is the key idea of this paper, which is to obtain the new space-charge-influenced Twiss parameters to modify the RDTs. 
	
	Here is an example on how to establish modified RDTs compensation. When a high-intensity coasting beam is injected into a HIAF-BRing whose bare tune is (9.47, 9.43), the tune spread would cross 3rd-order resonance lines like $3Q_y=28$ and $Q_x+2Q_y=28$, and the incoherent error driven resonances driven by the original skew and normal sextupoles in the ring would excite non-core particles on resonances. In order to suppress those resonances, modified RDTs of those particles on resonances should be minimized. First of all, get $\beta_{sc,u}$ through RMS envelope equation and $\phi_{sc,u}$ through Eq.~(\ref{eq_sc_phase_advance}) with appropriate $\Delta\nu_{sc,u}$. The first group of modified RDTs, $f_{0030}(\beta _{sc,u},\phi_{sc,u})$, along the ring should be obtained by Eq.~(\ref{eq_modifiedRDTs}), where the new phase modulated by space charge is obtained by identifying $\Delta\nu_{sc,y}=9.333-9.43$ in Eq.~(\ref{eq_sc_phase_advance}). The second group of modified RDTs, $f_{1020}(\beta _{sc,u},\phi_{sc,u})$, along the ring should be obtained similarly by identifying $\Delta\nu_{sc,x}=9.390-9.47$ and $\Delta\nu_{sc,y}=9.305-9.43$. Minimizing these two groups of modified RDTs with at least 2 normal and 2 skew sextupole correctors would mitigate the beam response to these resonances. 
	This compensation scheme does not consider the coherent resonances, which is another important issue although not discussed in this paper. 
	
	There are 3 potential issues in this approach. The first one is that, as the chosen $\Delta\nu_{sc,u}$ is close to 0, Eq.~(\ref{eq_sc_phase_advance}) describe large-amplitude particles, which means the ignored space-charge-induced nonlinear force would become no longer negligible and Eq.~(\ref{eq_sc_phase_advance}) would be imprecise. According to the following simulation, the space charge nonlinear force would make Eq.~(\ref{eq_sc_phase_advance}) imprecise for halo particles \cite{halo_particles} with large amplitude at the edge of the beam, as the dynamics of halo particles is complicated for the strong nonlinearity at the edge of the beam. The second one is that, while the detuning of space charge nonlinear terms is considered, the space charge nonlinear effect and corresponding resonances which might be driven are not considered, which might raise issues on resonances that are driven at high order by space charge itself. The third one is that, chromaticity and synchrotron motion, leading to periodic resonance crossing, are not considered in modified RDTs. 
	More examples of using Eq.~(\ref{eq_modifiedRDTs}) for compensation can be found in Sec.\ref{section3}
	
	Measuring modified RDTs experimentally remains a significant challenge. Modified RDTs describe the amplitude-dependent nonlinear motion of each single particle in a high-intensity beam, which means that each single particle has its own incoherent transverse motion. In order to measure modified RDTs, the oscillation of each single particle should be obtained. However, conventional beam oscillation detectors, like BPMs, could only obtain information of the beam charge center. Developing a new approach to measure modified RDTs requires further investigation. If the measured real lattice model with error information is precise enough, modified RDTs can be directly calculated. So one possible scheme to experimentally apply modified RDTs compensation is to measure optics and errors as precise as possible to get the real lattice model, which is feasible but still challenging. In the future modified RDTs experiments of HIAF-BRing, it is planned to precisely measure the optics and multipole errors. Errors would be obtained by 2 approaches, measuring the magnetic field and the measuring difference between RDTs obtained by each pair of BPMs in low intensity case. Through comparing errors obtained by these 2 approaches, precise errors in the lattice might be obtained and modified RDTs compensation scheme could be applied. 
	
	Modified RDTs explain the nonlinear behavior of incoherent error driven resonances and facilitates the compensation for each groups of errors in simulations. Without modified RDTs, it might take days, even months, to find the compensation for a group of errors through scanning the correctors in simulations. With modified RDTs, it takes minutes, significantly simplifying the assessment of corrector parameters, like positions, maximum strengths and polarities. Besides, modified RDTs might fundamentally alter the principles of lattice design for a high-intensity machine. For example, during the design of a lattice, the phase advance between a pair of chromaticity sextupoles should be $\Delta \phi_u = \pi$ to suppress their nonlinearity. But according to modified RDTs, this design would cause nonlinearity in a high-intensity machine since the nonlinear behavior of each particle depends on the phase advance modulated by space charge. As a result, greater consideration must be given to modified RDTs in the lattice design of high-intensity machines.

	\section{\label{section3}Simulation}
	
	In this section, simulations are performed on the GPU version of CISP to reduce the computational time. The space charge model in all simulations is the sliced 2.5D PIC. The space charge field is numerically evaluated by a $128\times128$ grid in the transverse plane and 40 slices in the longitudinal plane through the Green's function with a free boundary. All elements along the ring are sliced with the maximum length of 0.1 m and the effect of the space charge on particles is treated as thin elements which are evaluated at the end of every sliced element.
	
	\subsection{\label{appendix_structure_resonance}Systematic resonance analysis}
\begin{figure}[htbp]
		\centering
		\centering  
		\subfigure[]{
			\includegraphics[width=4.1cm]{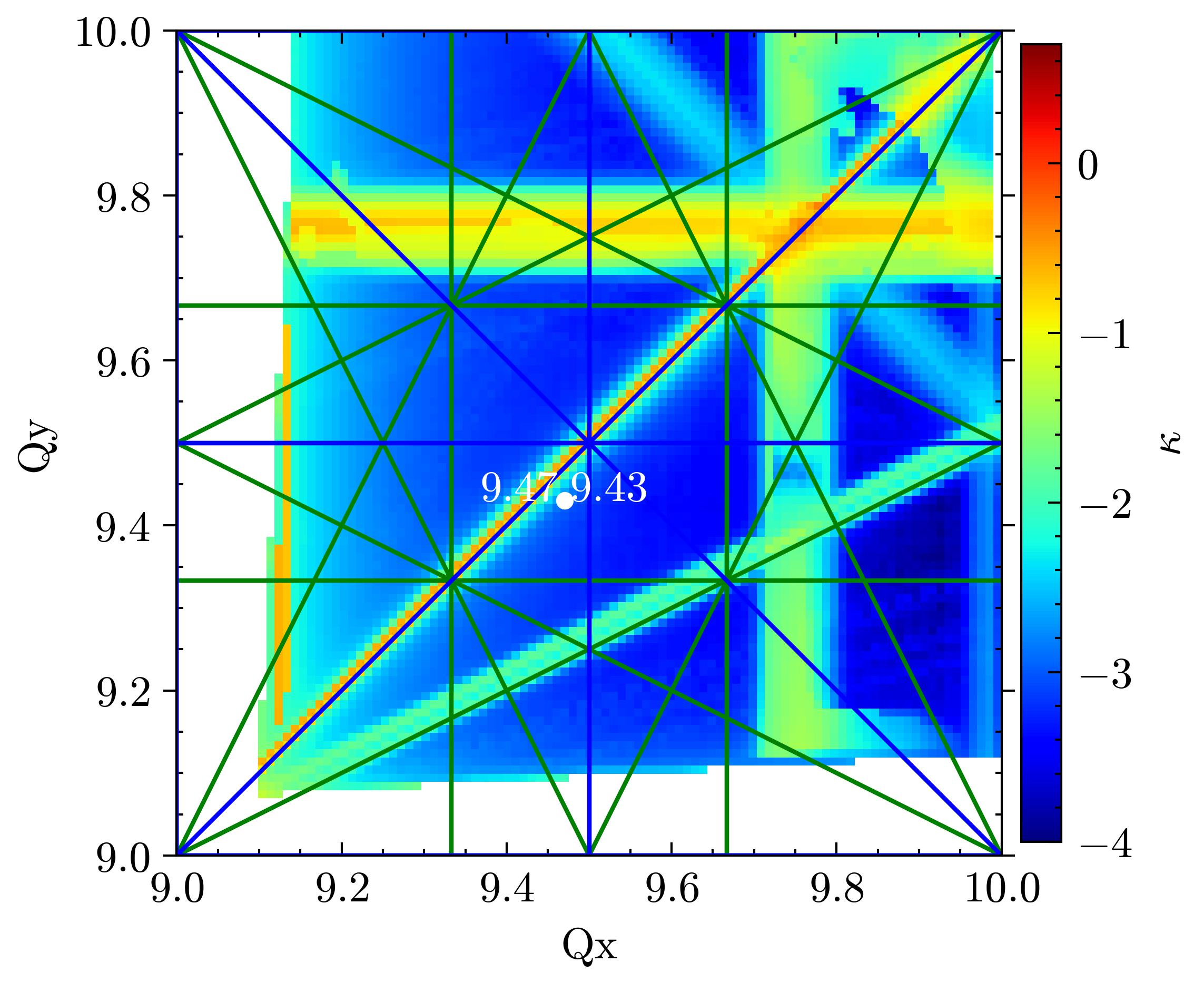}
		}
		\subfigure[]{
			\includegraphics[width=4.1cm]{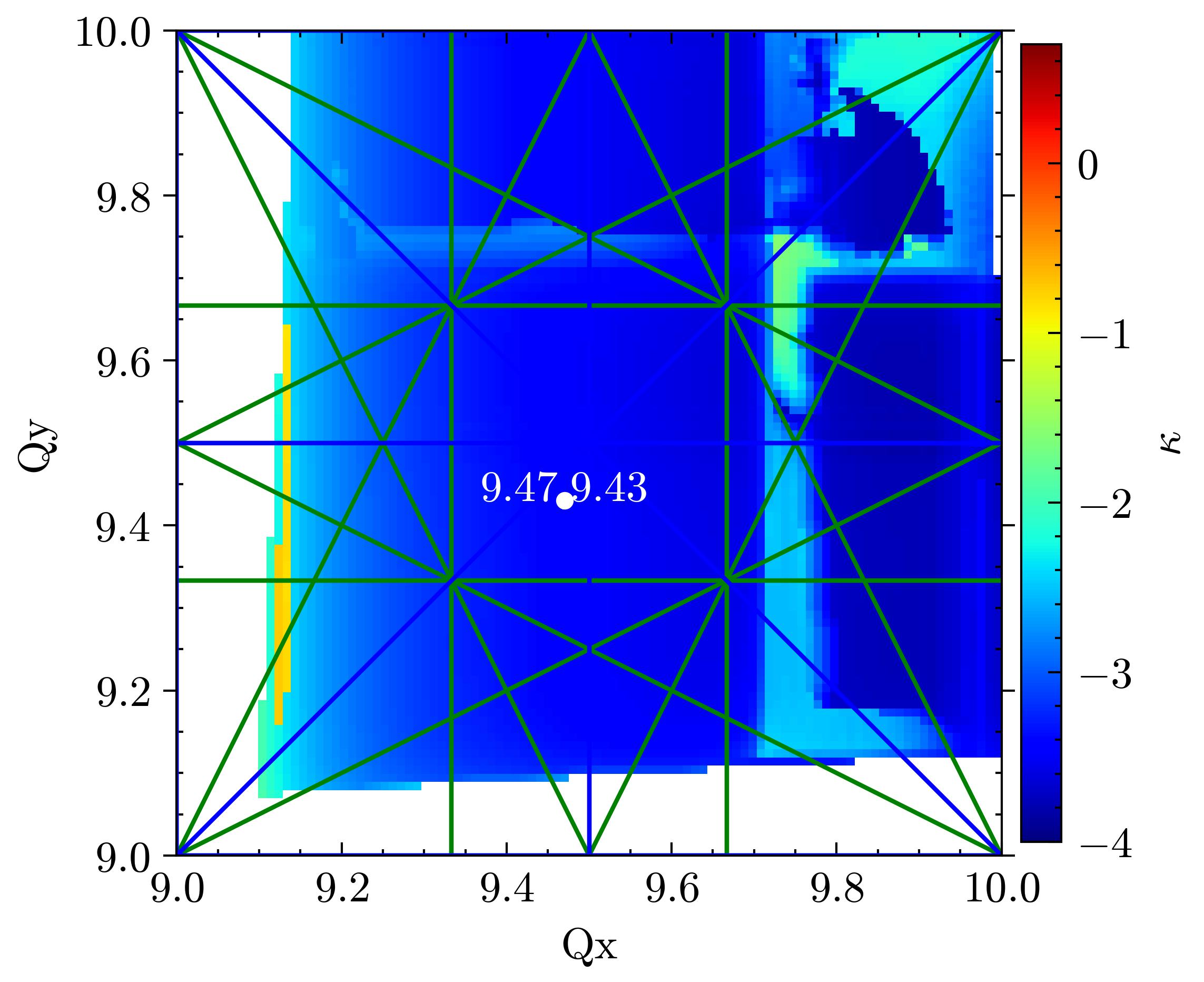}
		}
		\caption{
			\label{fig:systematic_resonance_analysis} 
			The figures show the bunched-beam simulation of HIAF-BRing tune scan map, where the white area has no data due to the unable tune matching. The max tune shift $\triangle\nu_y=-0.04$ in the (a) subgraph, and space charge is not included in the (b) subgraph. This simulation is constructed for 100 turns without magnetic field imperfections.
		}
	\end{figure}

	Nonlinear resonances are classified into systematic and error driven resonances. Systematic nonlinear resonances are located at $\ell = \mathrm {P} \times integer$, where $\mathrm {P}$ is the superperiod of an accelerator \cite{RN88}. A systematic resonance must be avoided, both in the simulation and in the reality. The variable $\kappa$, which describes the emittance change, is defined as:
	\begin{equation}
		\kappa = \log_{10}{(\frac{\left | \varepsilon _{rms,x1}- \varepsilon _{rms,x0}\right | }{\varepsilon _{rms,x0}}
			+\frac{\left | \varepsilon _{rms,y1}- \varepsilon _{rms,y0}\right | }{\varepsilon _{rms,y0}})} 
		\label{eq_Emit_change}
	\end{equation}
	where $\varepsilon _{rms,u0}$ presents the RMS emittance at turn 0, and $\varepsilon _{rms,u1}$ presents the RMS emittance at turn 100. The analysis of systematic resonance in the HIAF-BRing lattice is shown in Fig.~\ref{fig:systematic_resonance_analysis}. As the analysis shows, even without magnetic field imperfections, systematic resonances with space charge would be hit at $4Q_{x}=39$, $4Q_{y}=39$, $8Q_{x}=73$, $2Q_x+2Q_y=39$, and $-4Q_x+8Q_y=36$. Montague resonance driven by space charge would be hit at $2Q_{x}-2Q_{y}=0$ \cite{RN100}. Among these resonance lines, $4Q_{x}=39$, $4Q_{y}=39$, $8Q_{x}=73$ can be hit even without space charge. The bare tune of HIAF-BRing must locate away from these resonance lines.
	
	\subsection{Coasting-beam simulations\label{sec3_coasting}}

\begin{table}[htbp]
	\centering
	\caption{BRing parameters to test half-integer incoherent resonance} \label{half_integer_resonance}
	\begin{tabular}{@{}cc@{}}
		\toprule
		Parameters & Values \\ 
		\midrule
		Kinetic energy & 17MeV/nucleon \\ 
		Bare tune $(\nu_{x},\nu_{y})$ & (9.52,9.43) \\ 
		Intensity: ion num of $^{238}U^{35+}$  & $4.5\times10^{10}$ \\ 
		Simulation aperture & (300,150) $\pi$ mm mrad \\ 
		Injection RMS emittance $(\varepsilon_{x},\varepsilon_{y})$ & (42.10,14.06) $\pi$ mm mrad \\ 
		Transverse distribution & Gaussian truncated at $6\sigma$ \\
		Number of macroparticles & $1\times 10^6$ \\ 
		Max space charge tune shifts$(\Delta\nu_{x},\Delta\nu_{y})$ & (0.05,0.08)\\
		\bottomrule
	\end{tabular}
\end{table}
	
	This section presents coasting-beam simulations. The coasting beam is uniformly distributed in the longitudinal plane with minimal momentum spread in order to minimize the influence of longitudinal effects, such as the modulation of synchrotron motion on space charge effects and chromaticity. The primary objective of this paper, the feasibility of modified RDTs, would be validated in the simulations of this subsection. Two groups of simulations are presented in this subsection. The first group is constructed under resonance crossing at one single half-integer resonance lines. The second group is constructed under resonance crossing at 5 3rd-order resonance lines.
	
\begin{figure}[htbp]
		\centering
		\includegraphics[width=8.6cm]{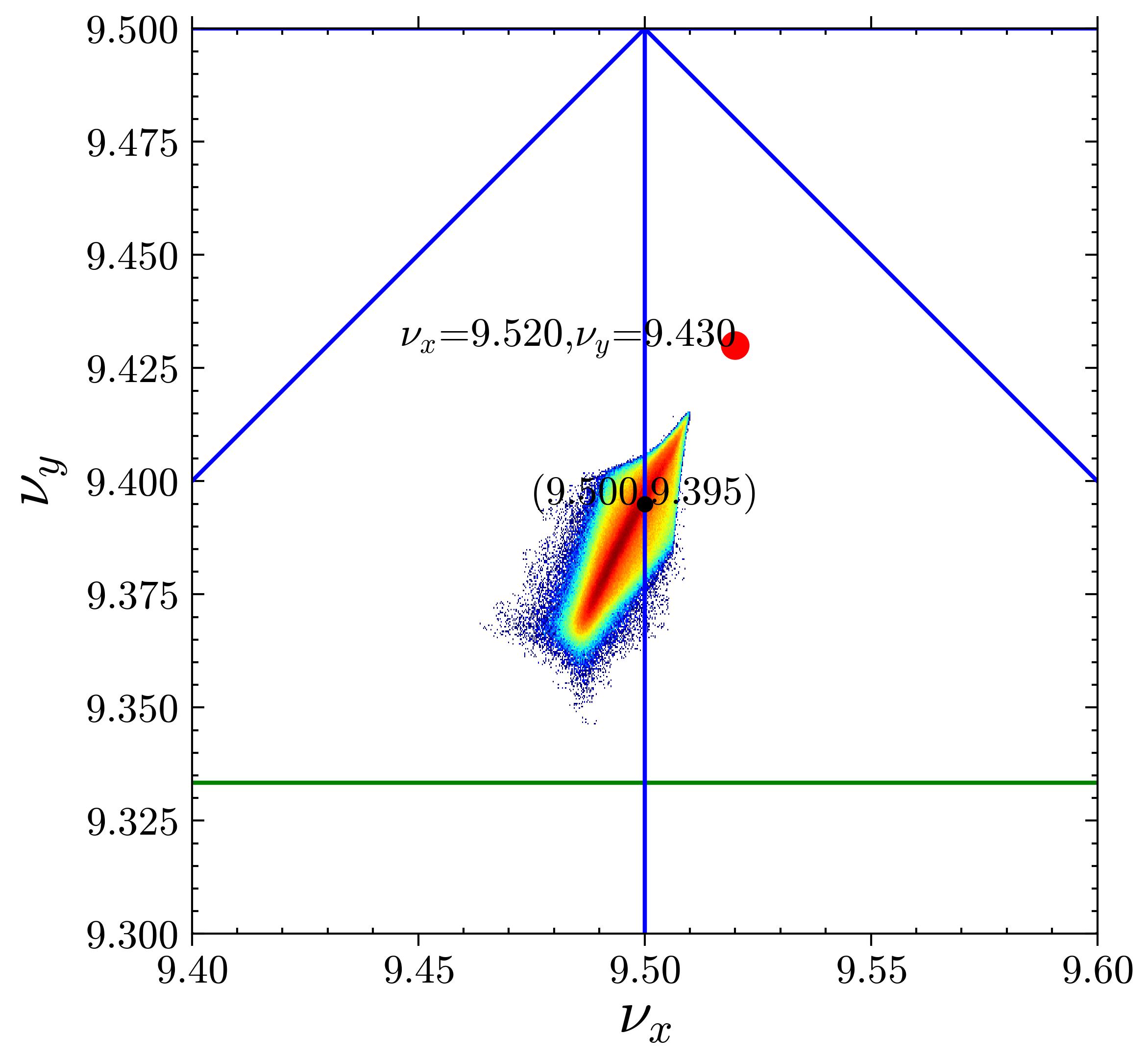}
		\caption{
			\label{fig_Half_integer_tune_spread_noC} 
			The space charge tune spread map using parameters in Table~\ref{half_integer_resonance}, under the resonance crossing at $2Q_{x}=19$. The black dot is one of the intersections of the tune spread and the resonance line, which could be used in Eq.~(\ref{eq_sc_phase_advance}). The tune footprint is acquired by evaluating the transverse phase advance of each particle over 1 turn.
		}
	\end{figure}

	The first group is to validate modified RDTs on half-integer incoherent error driven resonance $2Q_{x}=19$, which is chosen because it does not correspond to a systematic resonance, as illustrated in Fig.~\ref{fig:systematic_resonance_analysis}. To investigate the half-integer incoherent error driven resonance at $2Q_{x}=19$, the parameters of HIAF-BRing are adjusted as detailed in Table~\ref{half_integer_resonance}. The corresponding tune spread map is shown in Fig.~\ref{fig_Half_integer_tune_spread_noC}, where the core particles are free of resonances and non-core particles are excited by the incoherent error driven resonance. In later simulation, quadrupole errors, which come from Gaussian random values, $3\sigma(\Delta B_1/B_1) = 1\times10^{-3}$, on all quadrupoles, would be introduced in order to excite the half-integer incoherent error driven resonance and corresponding compensation schemes using modified RDTs or regular RDTs are applied to mitigate the beam response to this resonance.
	
\begin{figure}[htbp]
		\centering
		\includegraphics[width=8.6cm]{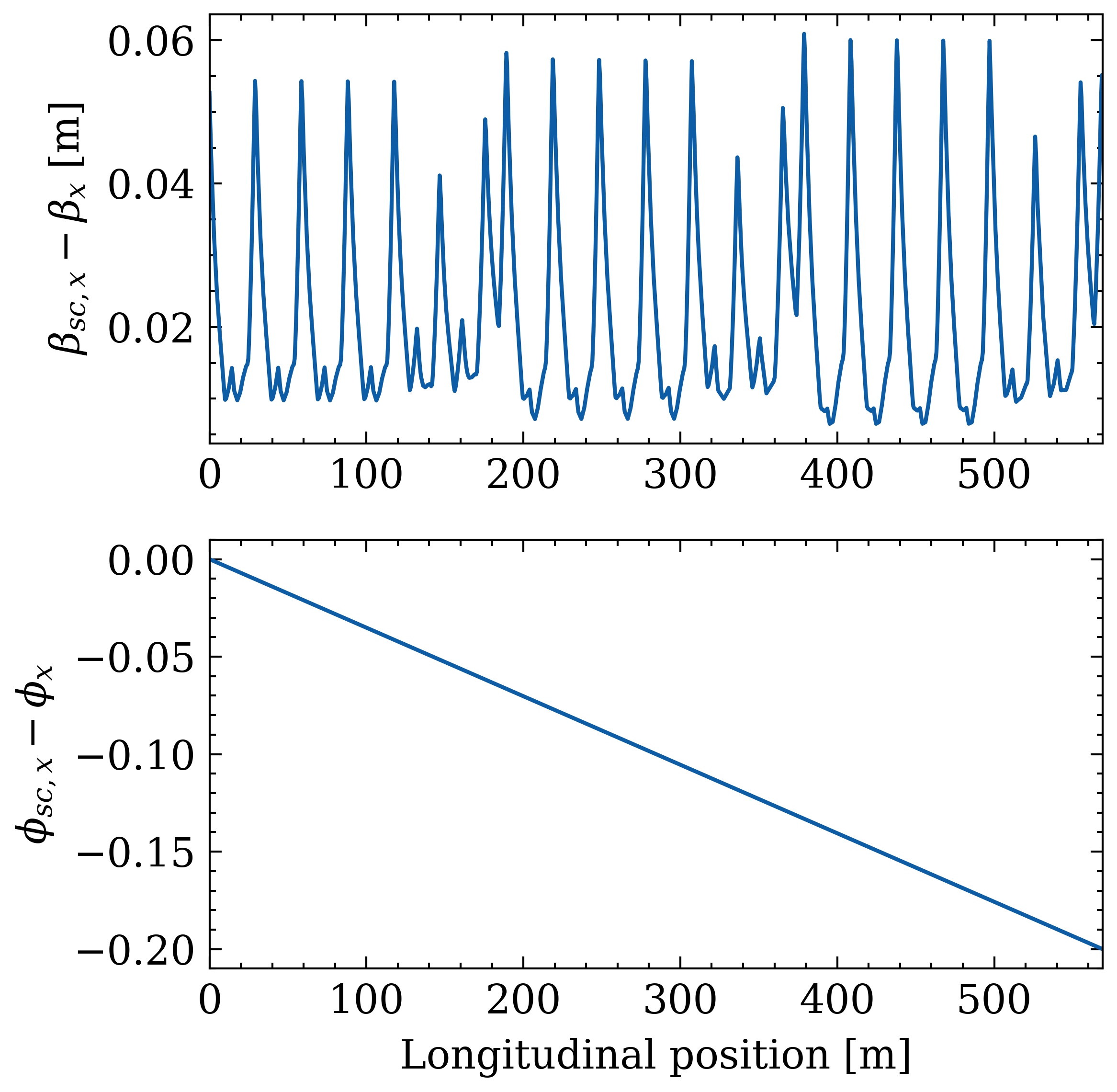}
		\caption{
			\label{fig_newTwiss} 
			This two subgraphs shows the horizontal Twiss differences between the new Twiss obtained by Eq.~(\ref{eq_Hill_tune_example}) and the original Twiss at (9.52,9.43).
		}
	\end{figure}

	For modified RDTs, since the tune spread would cross $Q_{x}=9.5$, substitute $\Delta\nu_{sc,x}=(9.5-9.52+o)$ ($o$ asymptotically approaches zero to prevent the denominator of $f_{jklm}^{(b)}$ from going to zero.) into Eq.~(\ref{eq_sc_phase_advance}):
	\begin{equation}
		\phi_{sc,u}(s) = \phi_{linear,u}(s)+\frac{s}{C} \times (-0.199)
		\label{eq_Hill_tune_example}
	\end{equation}
	where the additional 0.001 is the $o$. Then, get the corresponding $\beta_{sc,x}$ through envelope equation and $\phi_{sc,x}$ of space-charge-influenced particles which cross $Q_{x}=9.5$ through Eq.~(\ref{eq_Hill_tune_example}). As shown in the bottom subgraph of Fig.~\ref{fig_newTwiss}, the tune obtained by Eq.~(\ref{eq_Hill_tune_example}) approx 9.5. Finally, use the new Twiss ,$\beta_{sc,x},\phi_{sc,x}$, and Eq.~(\ref{eq_modifiedRDTs}) to get modified RDTs along the ring:
	\begin{equation}
		f_{2000}(\beta_{sc,x},\phi_{sc,x})
		\label{eq_modifiedRDTs_example}
	\end{equation}
	The whole progress is recorded as
	\begin{equation}\label{eq_half_integer_compensation}
		\Delta\nu_{sc,x}=9.501-9.520
		\to f_{2000}(\beta _{sc,x},\phi_{sc,x})
	\end{equation}
	Minimizing modified RDTs through additional quadrupole correctors would achieve the compensation. The method to minimize the RDTs has been well established in the previous study. In a word, minimizing RDTs is to introduce additional multipoles into the lattice with magnetic correctors for mitigating the original RDTs driven by corresponding multipoles due to various imperfections, including but not limit to alignment errors, rotation errors, and multipole errors in the magnet itself. Details on minimizing RDTs could be found in \cite{RN31,RN89}.
	
	For regular RDTs, first get the Twiss parameters at (9.52,9.43) from MAD-X \cite{RN99}, $\beta_{x}(9.52,9.43),\phi_{x}(9.52,9.43)$. Then, get RDTs directly through this Twiss and minimize the RDTs along the ring.
	\begin{equation}
		f_{2000}(\beta_{x}(9.52,9.43),\phi_{x}(9.52,9.43))
	\end{equation}
	Regular RDTs can also be obtained by PTC\_NORMAL command in MAD-X.
	
	The difference among these 2 schemes is the Twiss: modified RDTs use space-charge-influenced Twiss, regular RDTs use the Twiss at operating point (9.52,9.43). According to Fig.~\ref{fig_newTwiss} and the max incoherent tune shift $\Delta\nu_{sc,x}=-0.04$, the difference of beta function between modified RDTs and regular RDTs is only $(\beta_{sc,x}-\beta_x)/\beta_x\approx0.004$, which is extremely small. The key difference is the phase. Due to the difference of the phase, modified RDTs and regular RDTs caused by quadrupole errors are quite different in every position along the ring, and modified RDTs and regular RDTs provided by quadrupole correctors are also different for the same reason. As a result, different permutations of quadrupole correctors' strength would be obtained when minimizing modified RDTs or regular RDTs. 
	
	\begin{figure}[htbp]
		\centering
		\includegraphics[width=8.6cm]{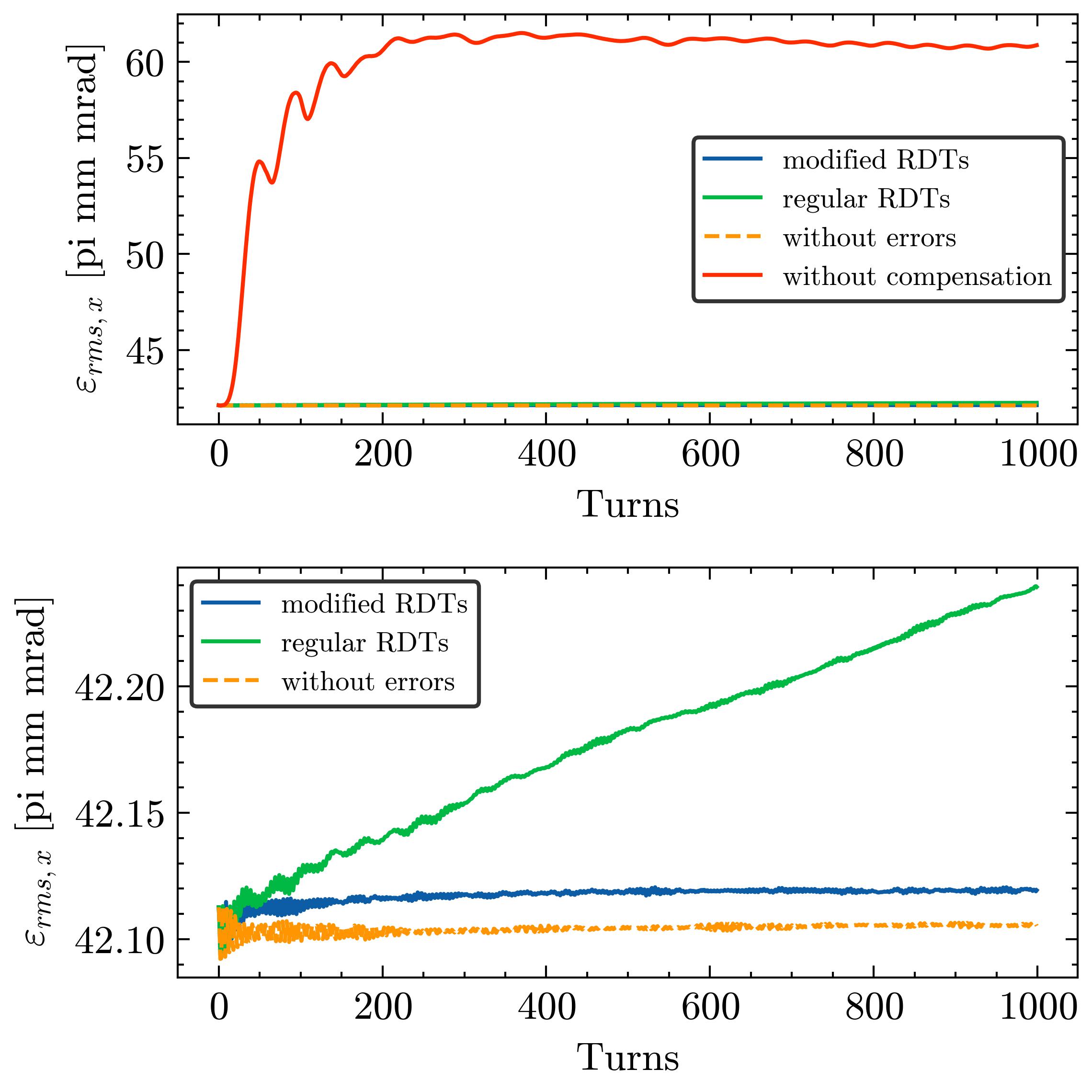}
		\caption{
			\label{fig_half_integer_emittance} 
			The horizontal emittance evolution under the tune spread in Fig.~\ref{fig_Half_integer_tune_spread_noC}. In the top subgraph, the emittance without compensation grows too much, causing that rest results almost overlap. The bottom subgraph shows more details about the comparison between these compensation schemes.
		}
	\end{figure}
	
	Figure.~\ref{fig_half_integer_emittance} shows the emittance evolution under 4 situations. The 'without errors' means that the simulation uses ideal lattice without any quadrupole error. The 'without compensation' means that the simulation includes quadrupole errors but applies no compensation scheme. 'modified RDTs' or 'regular RDTs' means the quadrupole errors are compensated through a group of quadrupole correctors whose strengths are determined by the scheme to minimize modified RDTs or regular RDTs. Rapid emittance growth is observed during the first 50 turns in the simulation without compensation. The rate of emittance growth decreases between 50 and 200 turns due to particle losses at the simulation aperture. Between 200 and 1000 turns, emittance growth stabilizes as the beam is no longer excited by the half-integer resonance, owing to the reduced tune spread caused by beam loss and emittance growth. Modified RDTs compensation scheme performs well, demonstrating small emittance growth during the first 30 turns and negligible growth in the rest of the evolution. In contrast, regular RDTs compensation scheme could also suppres the serious emittance growth without compensation, but the emittance increases by 0.3\% over 1000-turns and continues to rise based on its evolution trend.

	\begin{figure}[htbp]
		\centering
		\centering  
		\subfigure[]{
			\includegraphics[width=4.1cm]{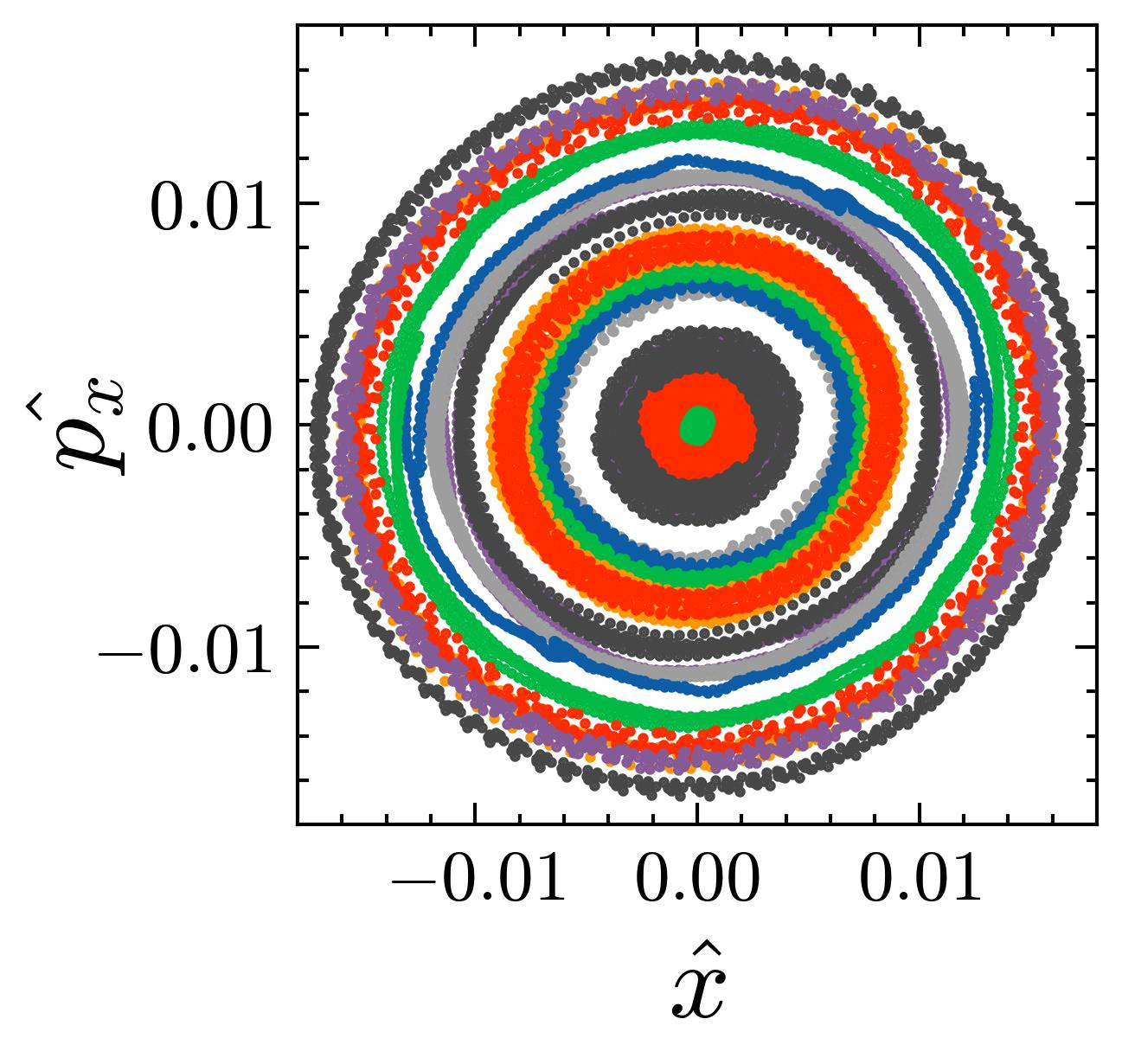}
		}
		\subfigure[]{
			\includegraphics[width=4.1cm]{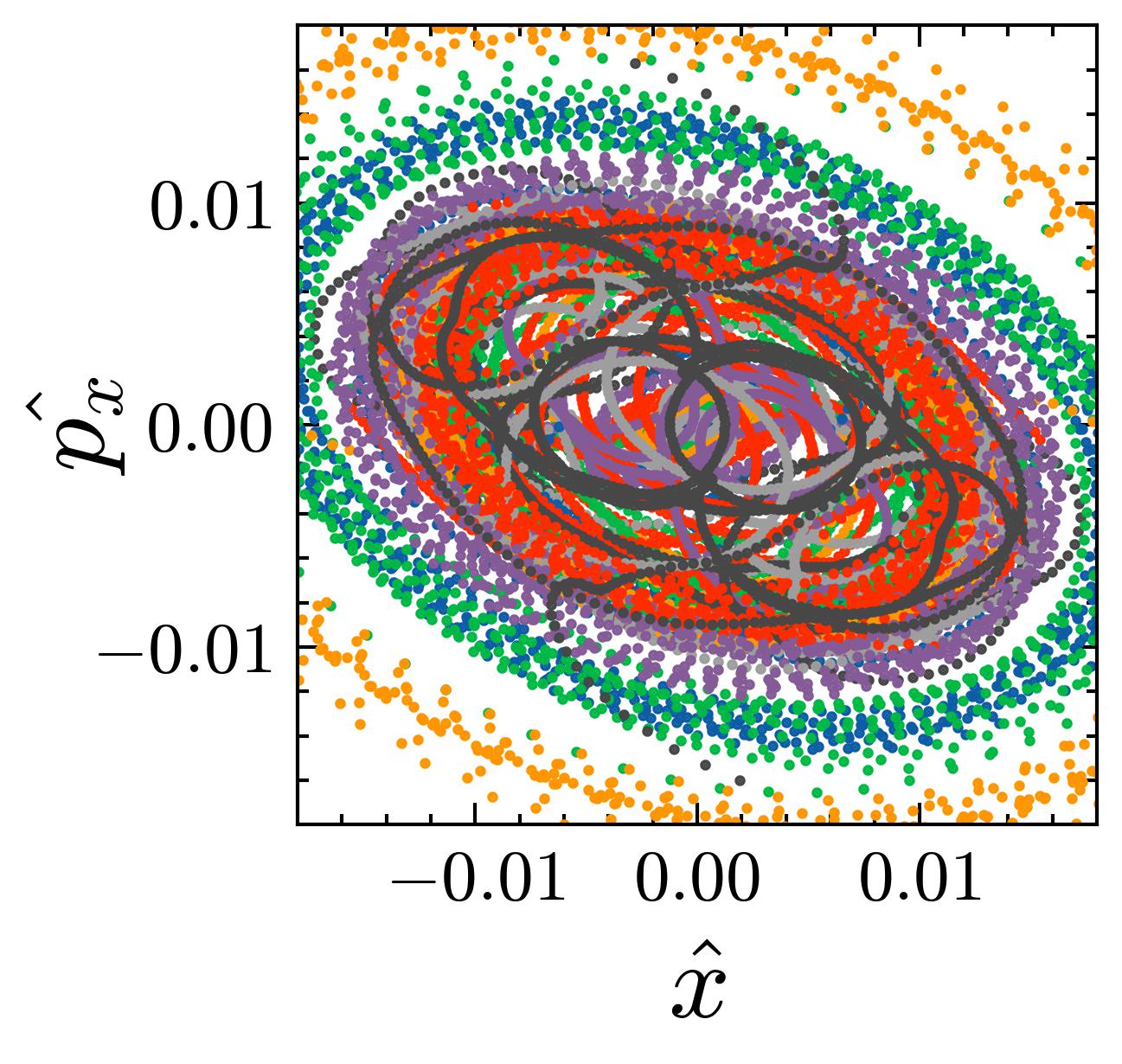}
		}
		\\
		\subfigure[]{
			\includegraphics[width=4.1cm]{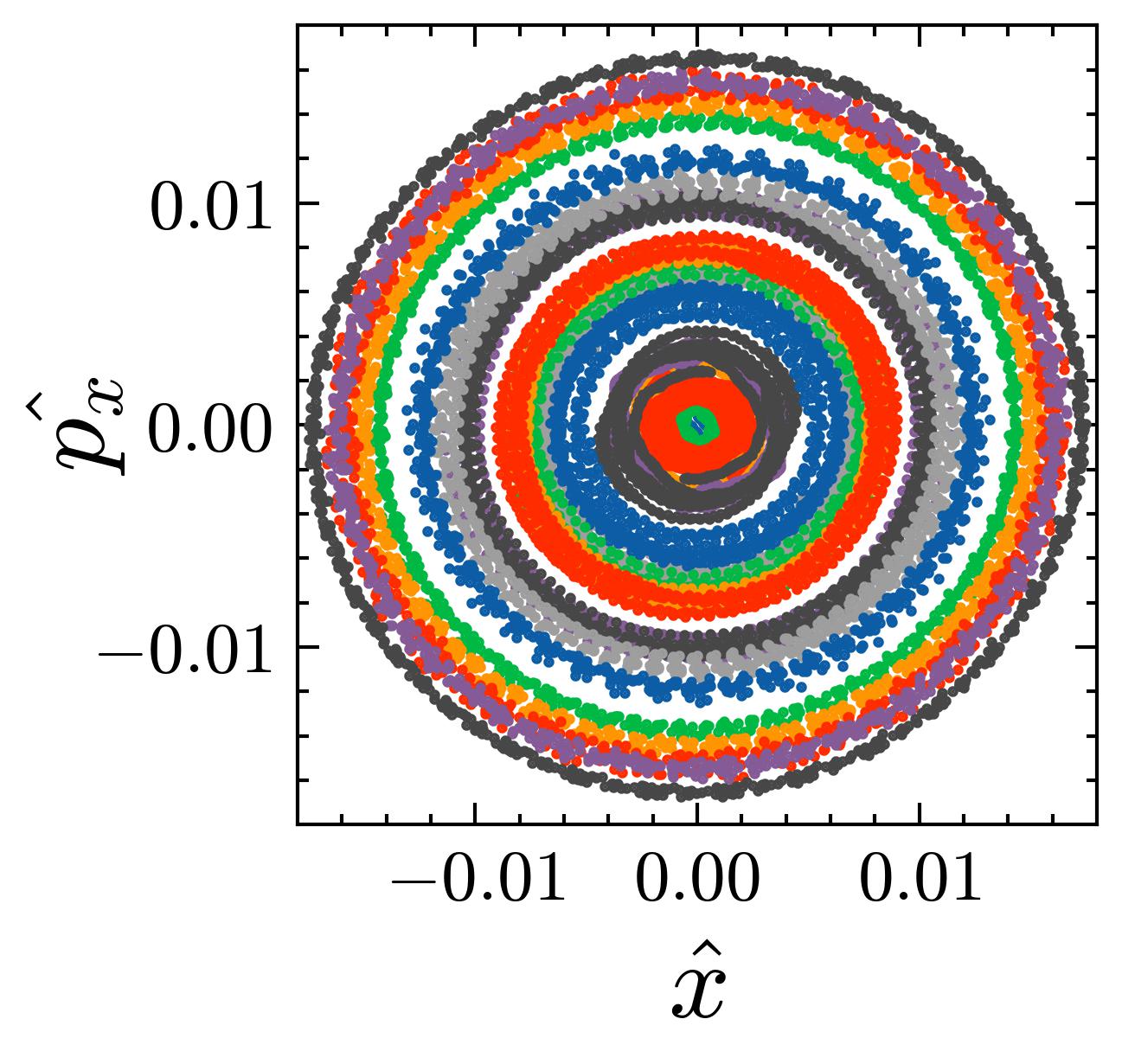}
		}
		\subfigure[]{
			\includegraphics[width=4.1cm]{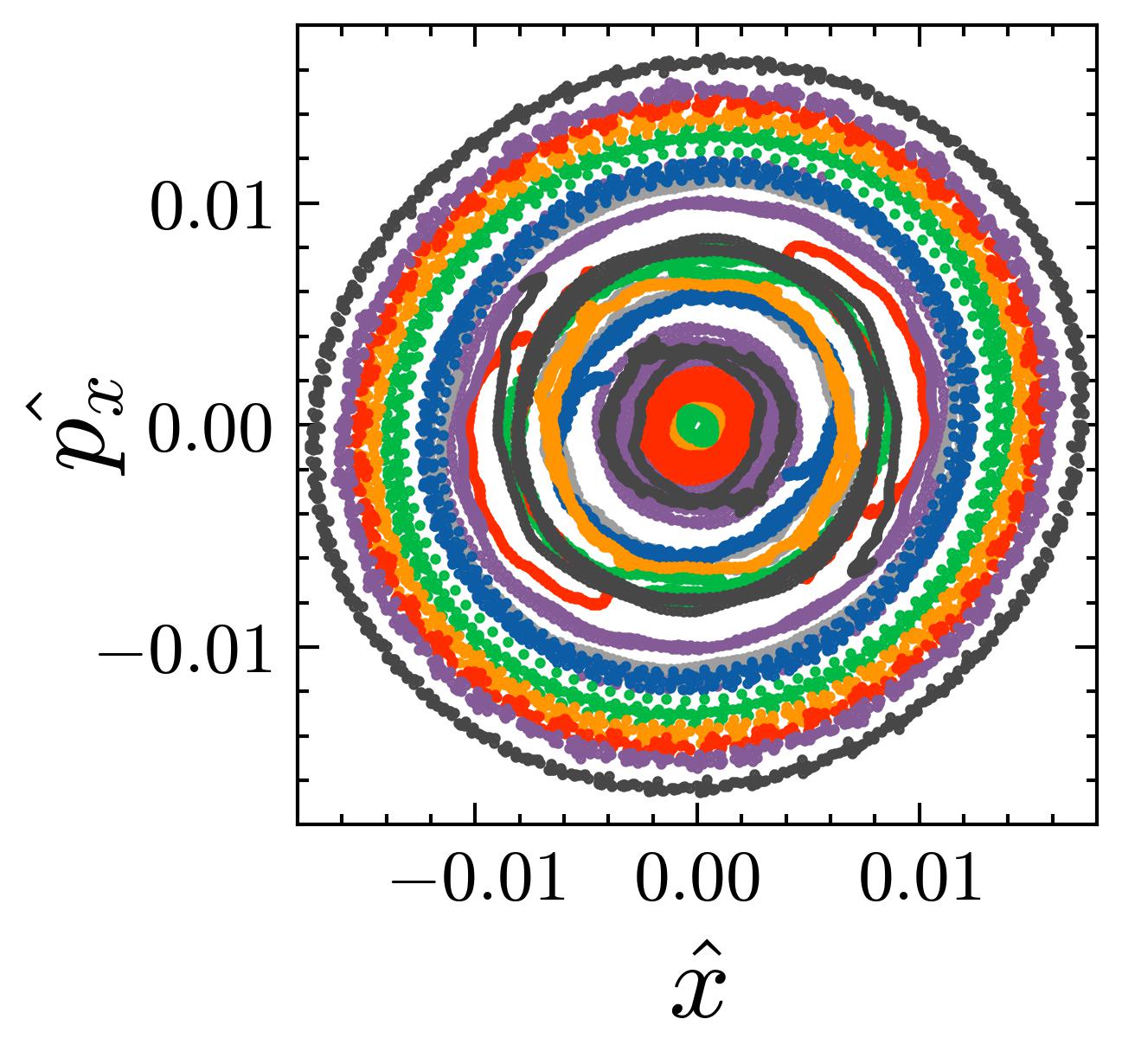}
		}
		\caption{
			\label{fig_half_integer_phase_space} 
			The particle track in normalized phase spaces under the tune spread in Fig.~\ref{fig_Half_integer_tune_spread_noC}. The (a) subgraph shows the the particle track in normalized phase space without errors. The (b) subgraph shows that without compensation. The (c) subgraph shows that with modified RDTs. The (d) subgraph shows that with regular RDTs. 
		}
	\end{figure}
	
	Figure.~\ref{fig_half_integer_phase_space} presents the particle track in normalized phase spaces. The initial 6D coordinates of particles with identical IDs are consistent across all four simulations. In the simulation without errors, the track of particles from small amplitude to large amplitude is recorded, along with their corresponding IDs. According to these IDs, the same particles are recorded in the rest of simulations. The simulation without compensation demonstrates the serious unstable particles motion in the phase space. The unstable motion is mitigated by the compensation with modified RDTs and regular RDTs. The compensation with regular RDTs does not have a perfect performance due to the insufficient compensation, resulting in resonance islands in the normalized phase space. Meanwhile, the track with modified RDTs shows little resonance behavior. It can be concluded that the compensation scheme with modified RDTs successfully mitigate the beam response to the half-integer incoherent error driven resonance.
	
	\begin{figure}[htbp]
		\centering
		\includegraphics[width=8.6cm]{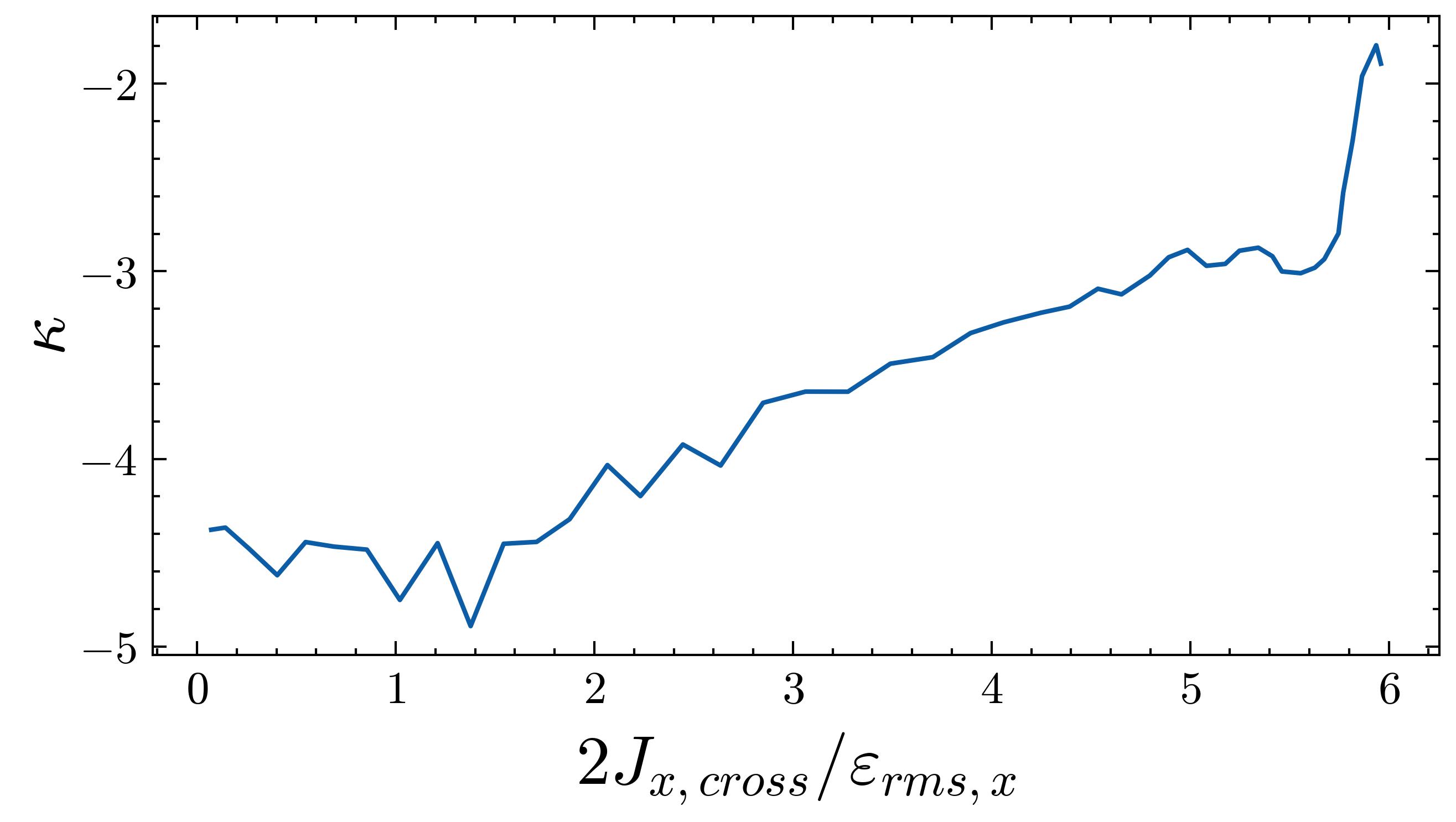}
		\caption{
			\label{fig_J_x/emit_vs_kappa} 
			The influence of the resonance crossing particles' amplitude on the effectiveness of compensation by modified RDTs. $\kappa$ is defined in Eq.~(\ref{eq_Emit_change}), and $\kappa$ here only describes x plane for 100 turns. $J_{x,cross}$ presents the action of particles located on the resonance line.
		}
	\end{figure}
	
	In order to investigate the influence of the resonance crossing particles' amplitude on the effectiveness of compensation by modified RDTs, an additional investigation is demonstrated in Fig.~(\ref{fig_J_x/emit_vs_kappa}). In this investigation, half-integer resonance crossing with quadrupole errors and corresponding compensation by modified RDTs are applied. Through adjusting the particle number presented by each microparticle, the entire space charge can be adjusted. As a result, the amplitude of resonance crossing particles can be changed while the beam distribution is unchanged. It can be observed in Fig.~(\ref{fig_J_x/emit_vs_kappa}) that the effectiveness of compensation by modified RDTs decreases as the amplitude of resonance crossing particles increases. In the area of [0, 2], modified RDTs have a perfect performance and such low emittance growth can be statistical errors. In the area of (2, 5.6], modified RDTs have a good but imperfect performance. In the area of (5.6, 6], which is the edge of the beam as the transverse distribution is Gaussian truncated at $6\sigma$, modified RDTs do not have a good performance. It can be concluded that, for Gaussian distribution, modified RDTs are perfect in [0, 2] $\sigma$, good but imperfect in (2, 5.6] $\sigma$, and not good in (5.6, 6] $\sigma$. 
	
	The second group of simulations is to validate modified RDTs on 3rd-order incoherent error driven resonances at $3Q_y=28$, $2Q_x+Q_y=28$ and $Q_x+2Q_y=28$. It is critical to investigate these resonances as they are major limitations for achieving high-intensity targets in the HIAF-BRing design. 
	
\begin{figure}[htbp]
		\centering
		\includegraphics[width=8.6cm]{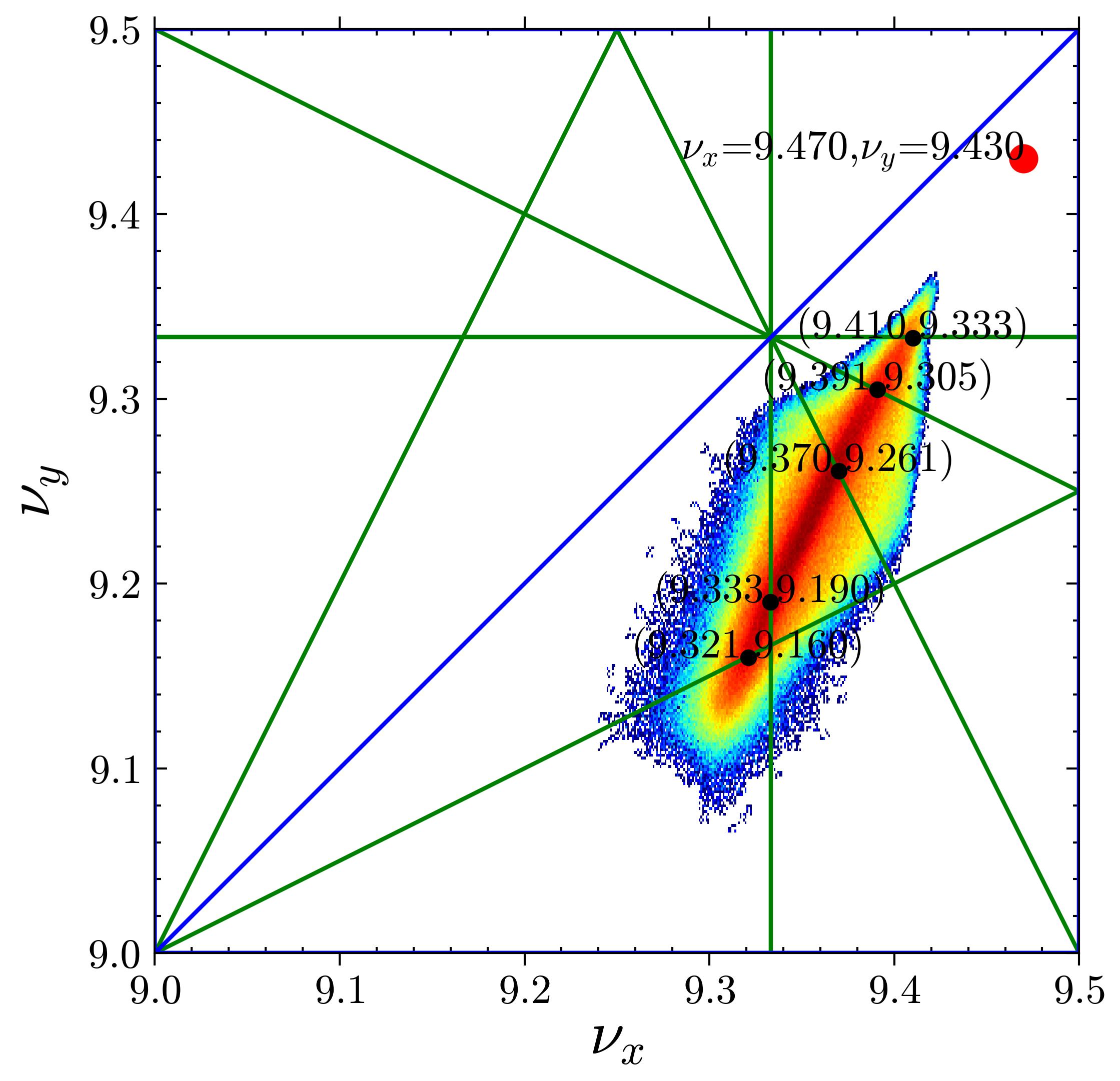}
		\caption{
			\label{fig_3rd_tune_spread_coasting} 
			The space charge tune spread map using parameters in Table~\ref{3rd_order_resonance}, under the resonance crossing at $3Q_x=28,Q_x+2Q_y=28,Q_x-2Q_y=-9$ driven by sextupole errors and $3Q_y=28,2Q_x+Q_y=28$ driven by skew sextupole errors. The black dots are intersections of the tune spread and resonance lines, which could be used in Eq.~(\ref{eq_sc_phase_advance}). This tune footprint is acquired by evaluating the transverse phase advance of each particle over 1 turn.
		}
	\end{figure}

	The parameters of HIAF-BRing are detailed in Table~\ref{3rd_order_resonance}. The intensity exceeds design value in order to ensure that the 3rd-order incoherent error driven resonances could be excited in coasting-beam case. The corresponding tune spread map is demonstrated in Fig.~\ref{fig_3rd_tune_spread_coasting}, where core particles are mainly affected by coherent resonances, non-core particles are additionally affected by incoherent resonances. In a simulation of full acceleration, where the beam transitions from a coasting beam to being captured and accelerated from 17 to 835 MeV/nucleon, the maximum space-charge-induced incoherent tune spread of the bunched beam closely matches that in Fig.~\ref{fig_3rd_tune_spread_coasting}, indicating that the space-charge-induced tune spread in this simulation is large enough to demonstrate expected incoherent resonances in HIAF-BRing. In later simulation, sextupole errors which come from Gaussian random values, $3\sigma(\Delta B_2/B_1) = 1\times10^{-3}$, on all quadrupoles, would be introduced in order to excite these 3rd-order incoherent error driven resonances and corresponding compensation with modified RDTs or regular RDTs would be applied to mitigate the beam response to these resonances. 
	
	\begin{table}[htbp]
		\centering
		\caption{BRing parameters to test 3rd-order resonance} \label{3rd_order_resonance}
			\begin{tabular}{@{}cc@{}}
				\toprule
				Parameters & Values \\ 
				\midrule
				Kinetic energy & 17MeV/nucleon \\ 
				Tune $(\nu_{x},\nu_{y})$ & (9.47,9.43) \\ 
				Intensity: ion num of $^{238}U^{35+}$  & $3.8\times10^{11}$ \\ 
				Simulation aperture & (300,150) $\pi$ mm mrad \\ 
				Injection RMS emittance $(\varepsilon_{x},\varepsilon_{y})$ & (42.10,14.06) $\pi$ mm mrad \\ 
				Transverse distribution & Gaussian truncated at $6\sigma$ \\
				Number of macroparticles & $1\times 10^6$ \\ 
				Max space charge tune shifts$(\Delta\nu_{x},\Delta\nu_{y})$ & (0.17,0.31)\\
				\bottomrule
			\end{tabular}
	\end{table}
	
	With the aperture set to the designed (200,100) $\pi$ mm mrad, 40\% of the beam is lost in the simulation of full acceleration when sextupole and skew sextupole errors are introduced. The unstable particle motion induced by these third-order resonances poses a significant challenge, limiting the achievable intensity in HIAF-BRing. During the simulation, it is observed that some non-core particles are still affected by incoherent resonances of $3Q_x=28$ and $Q_x-2Q_y=-9$. To enhance the compensation effectiveness, the compensation scheme for incoherent error driven resonances also includes these 2 resonance lines.
	
	For modified RDTs, 5 groups of modified RDTs should be obtained through the black dots by Eq.~(\ref{eq_modifiedRDTs})
	\begin{equation}\label{eq:wideeq}
		\begin{aligned}
			\Delta\nu_{sc,y}=9.333-9.43
			\to f_{0030}(\beta _{sc,u},\phi_{sc,u})
			\\
			\Delta\nu_{sc,x}=9.333-9.47
			\to f_{3000}(\beta _{sc,u},\phi_{sc,u})
			\\
			\left.\begin{matrix}
				\Delta\nu_{sc,x}=9.391-9.47
				\\
				\Delta\nu_{sc,y}=9.305-9.43
			\end{matrix}\right\}
			\to f_{1020}(\beta _{sc,u},\phi_{sc,u})
			\\
			\left.\begin{matrix}
				\Delta\nu_{sc,x}=9.370-9.47
				\\
				\Delta\nu_{sc,y}=9.261-9.43
			\end{matrix}\right\}
			\to f_{2010}(\beta _{sc,u},\phi_{sc,u})
			\\
			\left.\begin{matrix}
				\Delta\nu_{sc,x}=9.321-9.47
				\\
				\Delta\nu_{sc,y}=9.160-9.43
			\end{matrix}\right\}
			\to f_{1002}(\beta _{sc,u},\phi_{sc,u})
		\end{aligned}
	\end{equation}
	where the 5 groups of modified RDTs should be minimized by corresponding correctors. Note that compensating for one 3-order resonance would affect other 3-order resonances. So all sextupole correctors must be considered while minimizing each group of RDTs. A good approach to solve this issue is to compensate for all resonances simultaneously through expanding the equation (11) in the cited \cite{RN107} into the case of 5 resonances. Besides, the contribution of low-order RDTs on the high-order RDTs, called crossing terms \cite{RN117}, which would lead to residual high-order RDTs even though the low-order RDTs were well minimized, is another important mechanism. All high-order magnet fields in simulations of this paper are treated as thin multipoles, so this mechanism would not affect the simulation for now. But in the later resonance study on the real lattice, the crossing terms would also become vital.
	
	For regular RDTs, just get 5 groups of RDTs through the Twiss parameters at (9.47,9.43) and minimize them.
	
	\begin{figure}[htbp]
		\centering
		\includegraphics[width=8.6cm]{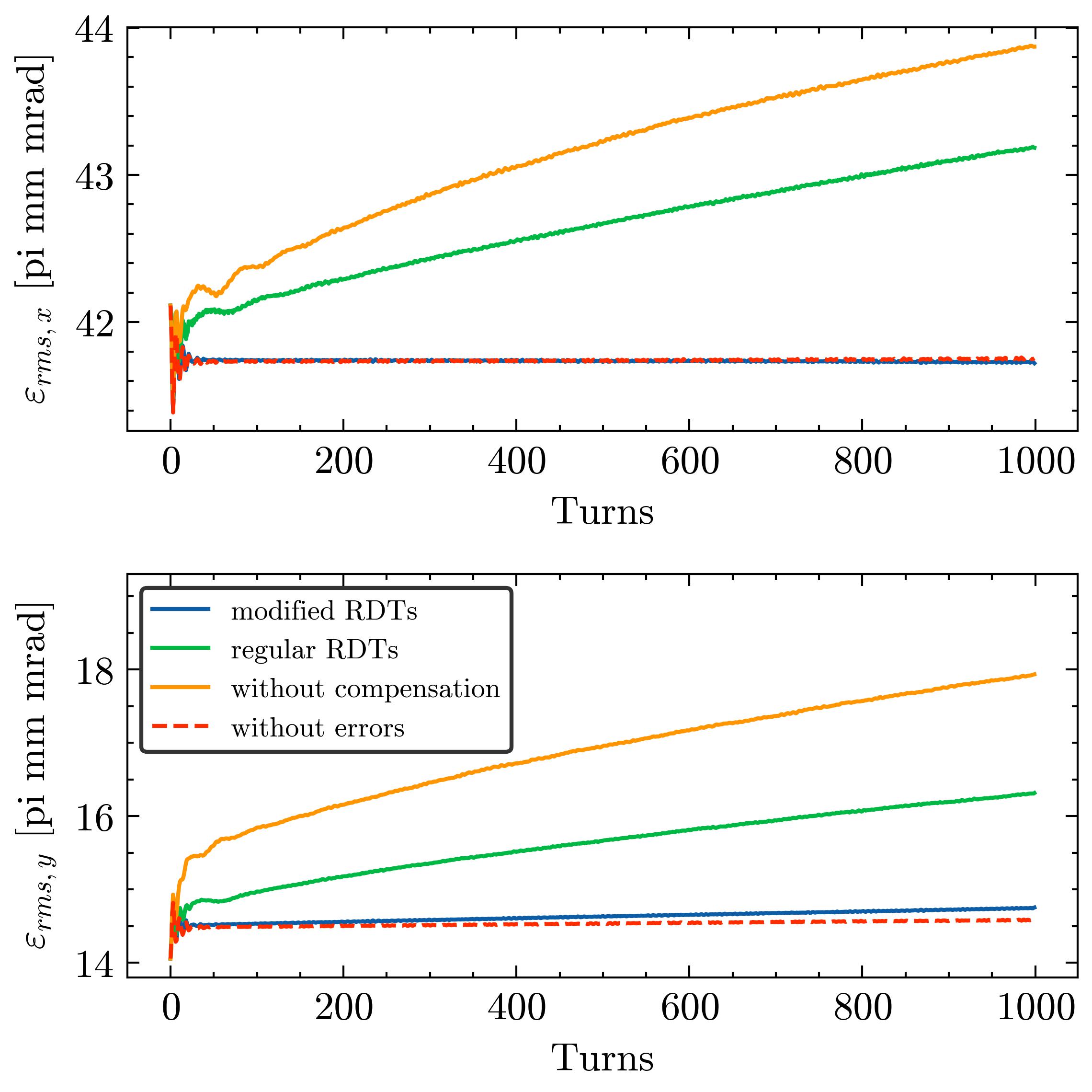}
		\caption{
			\label{fig_9.47_9.43_emit_chrom_off} 
			The emittance evolution under the tune spread in Fig.~\ref{fig_3rd_tune_spread_coasting}. The labels describe the same situation as that in Fig.~\ref{fig_half_integer_emittance}.
		}
	\end{figure}

    \begin{figure}[htbp]
		\centering  
		\subfigure[]{
			\includegraphics[width=4.1cm]{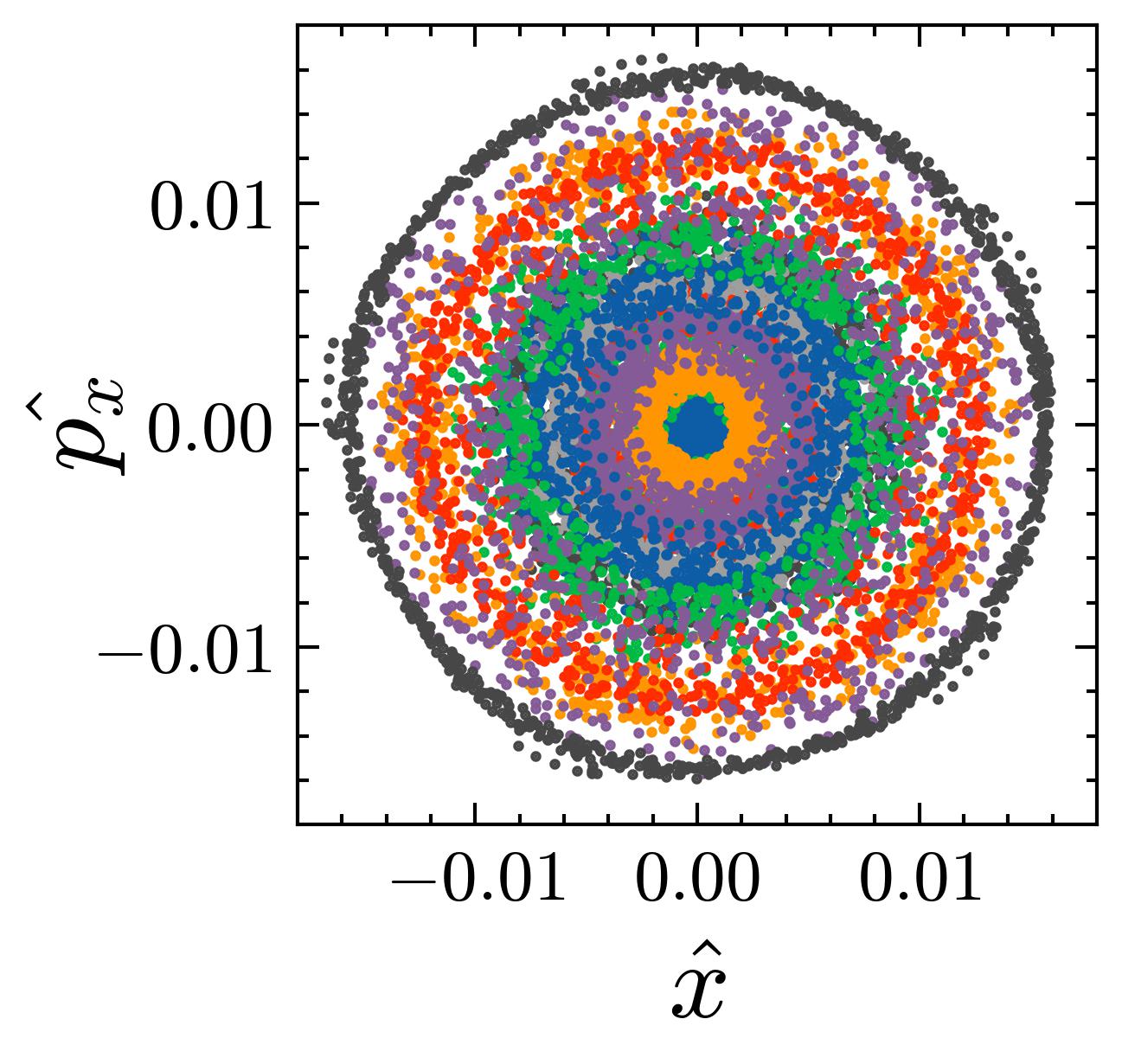}
		}
		\subfigure[]{
			\includegraphics[width=4.1cm]{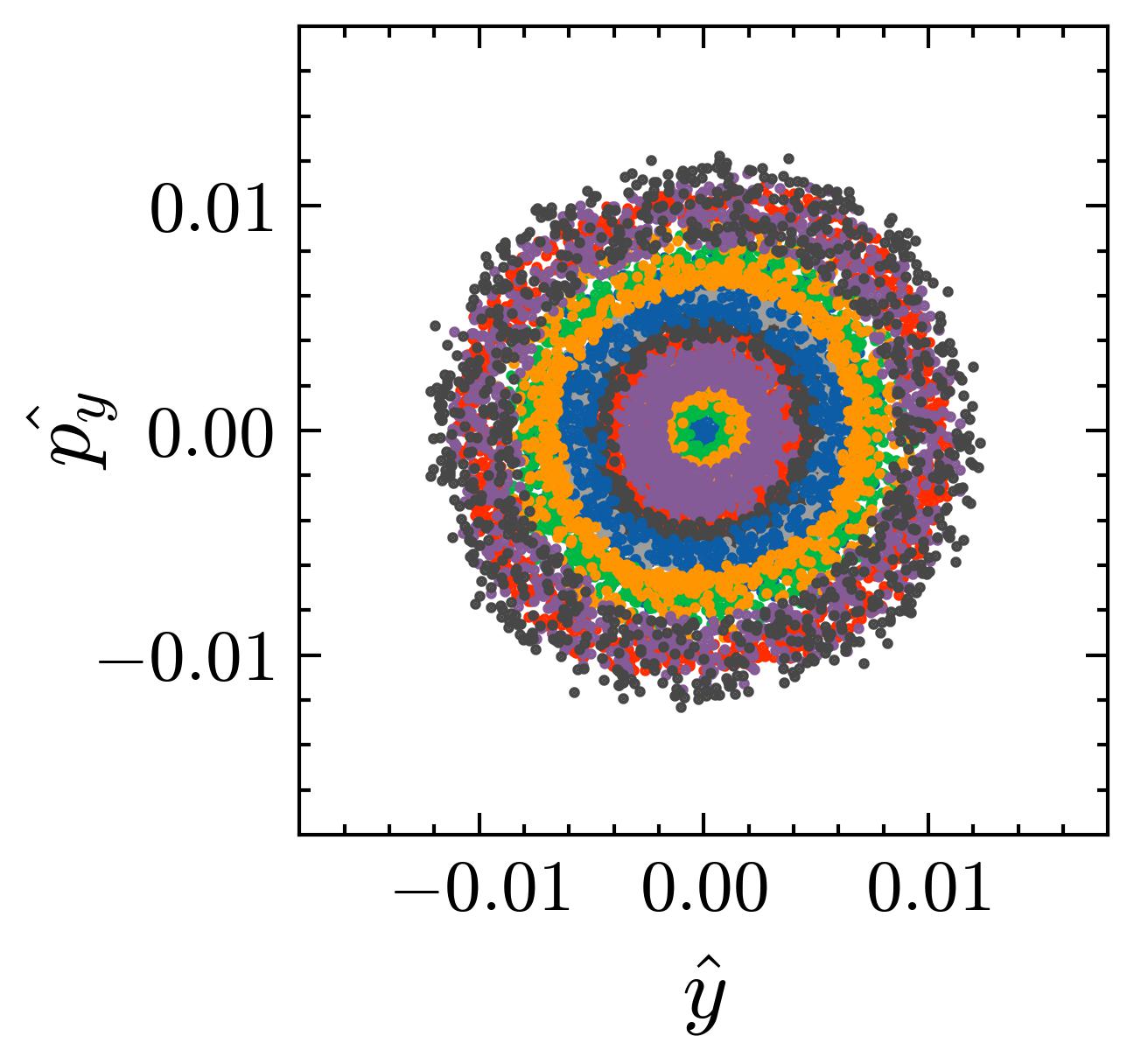}
		}
		\\
		\subfigure[]{
			\includegraphics[width=4.1cm]{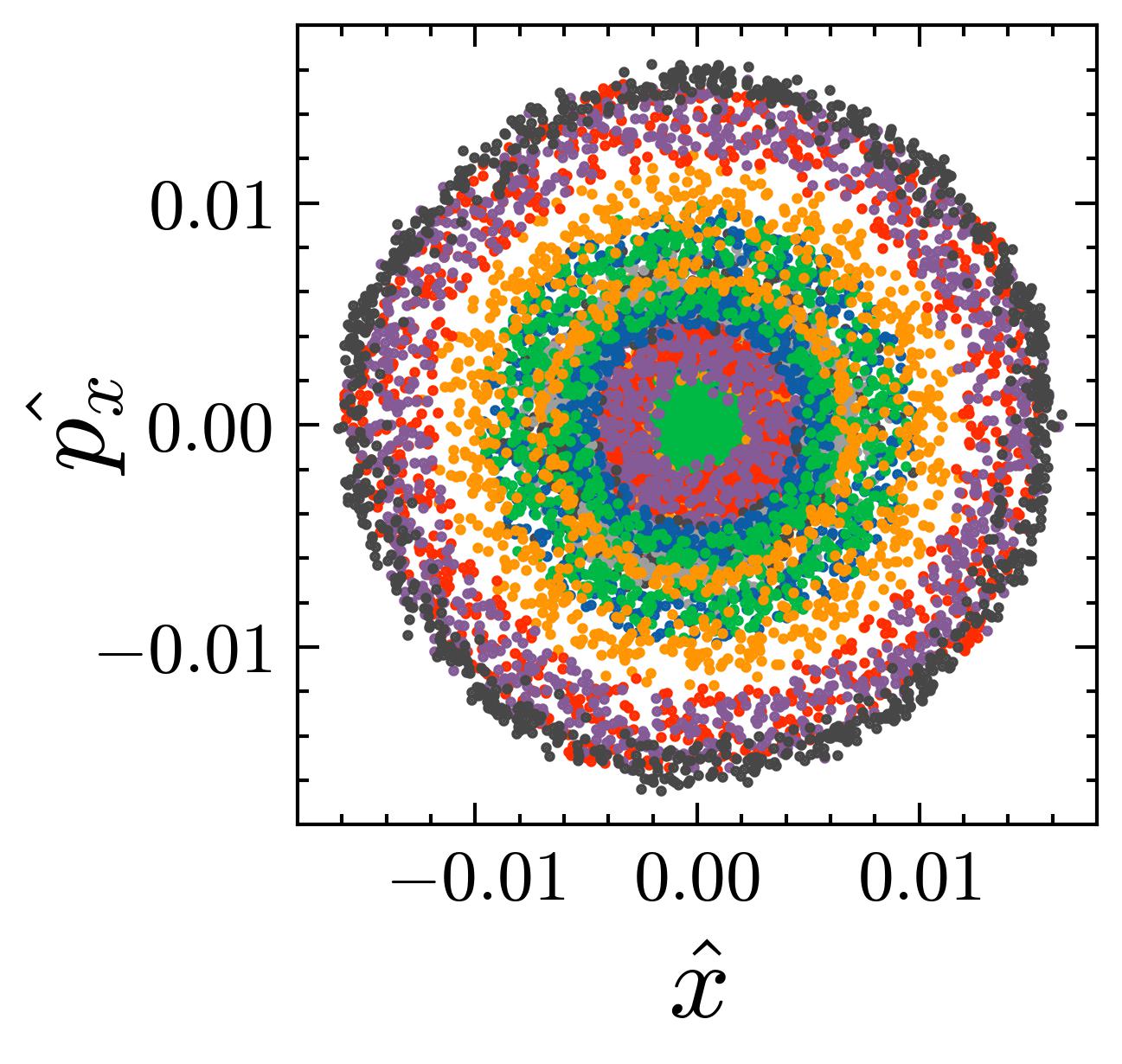}
		}
		\subfigure[]{
			\includegraphics[width=4.1cm]{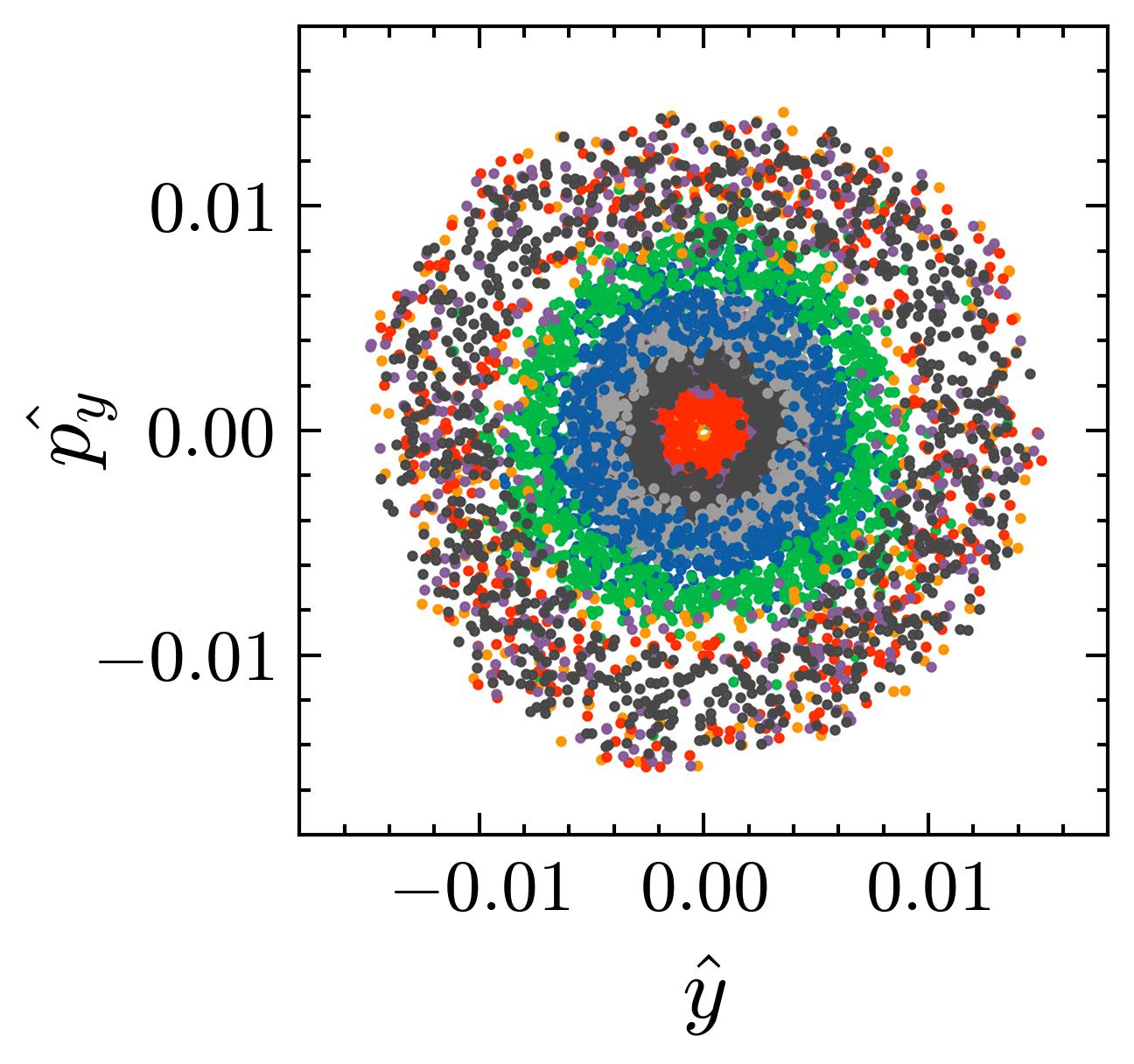}
		}
		\\
		\subfigure[]{
			\includegraphics[width=4.1cm]{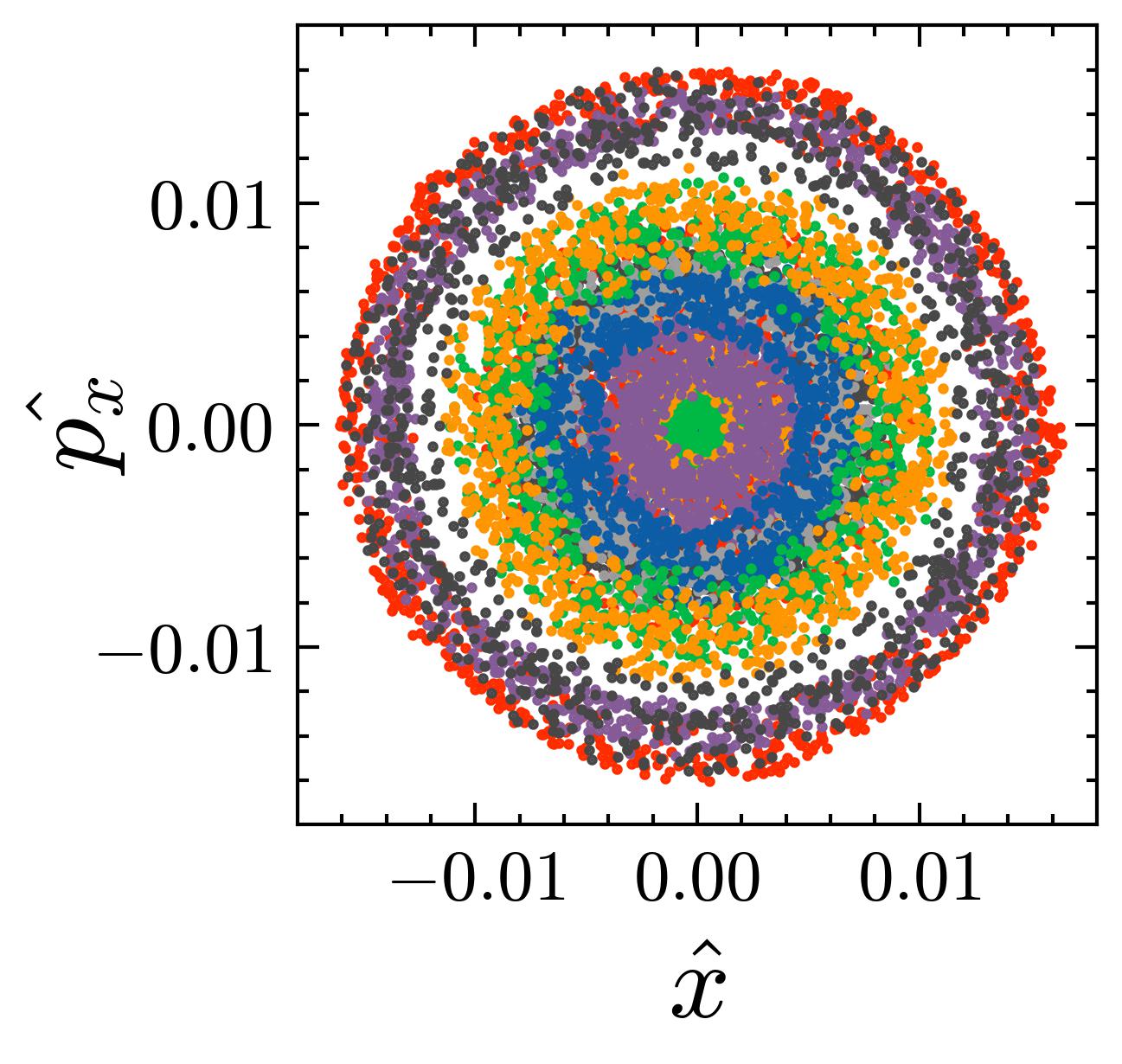}
		}
		\subfigure[]{
			\includegraphics[width=4.1cm]{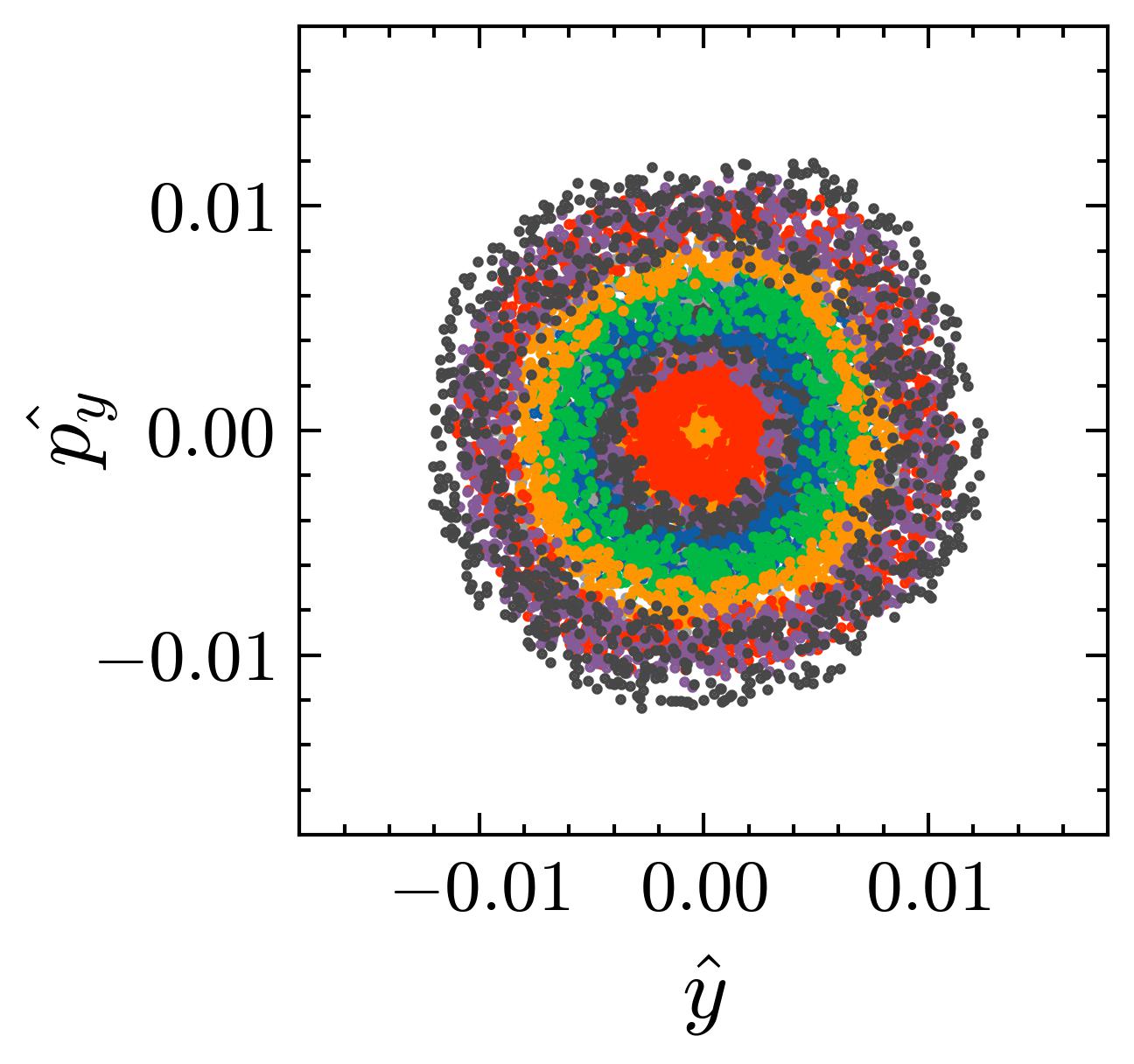}
		}
		\\
		\subfigure[]{
			\includegraphics[width=4.1cm]{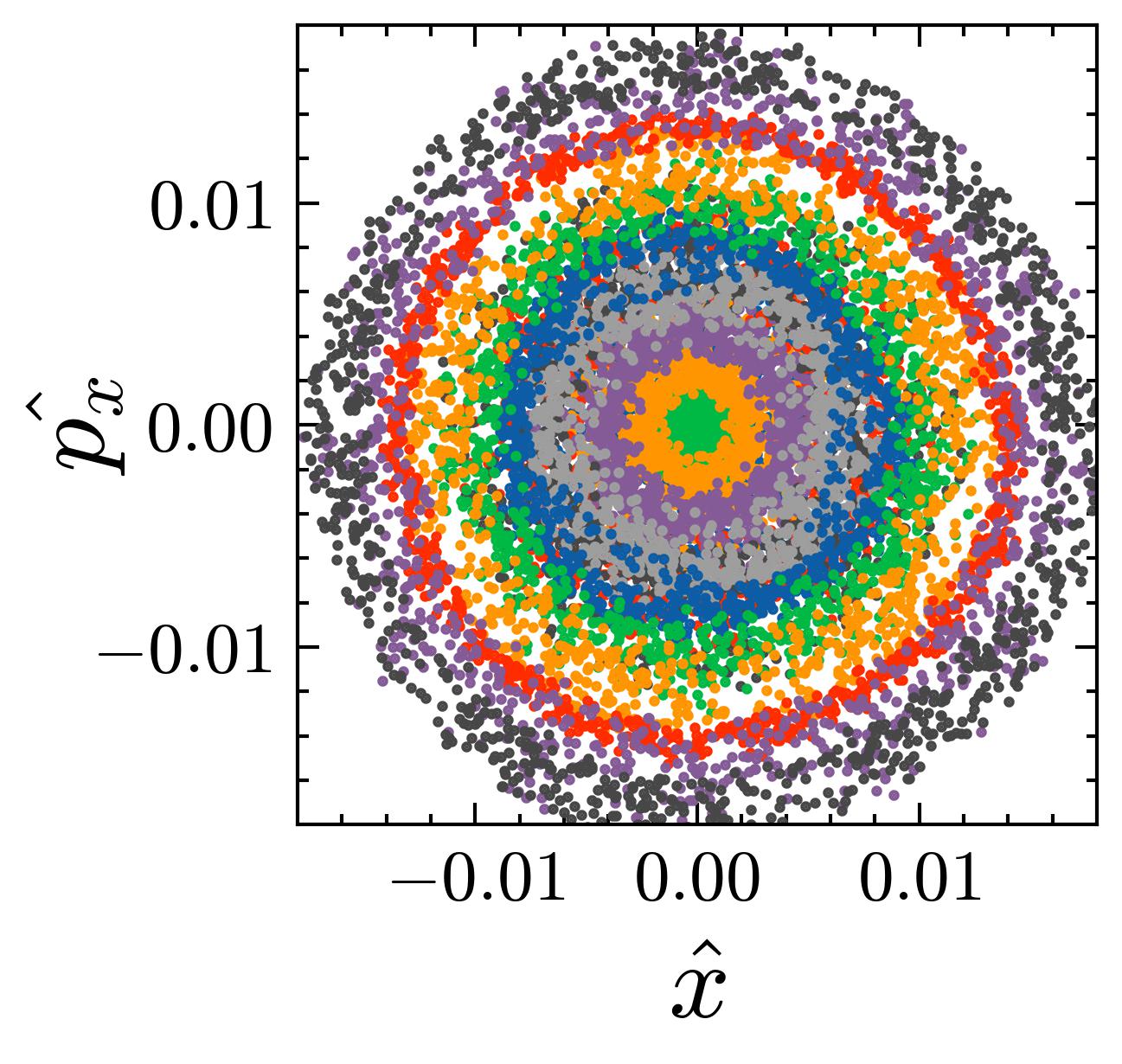}
		}
		\subfigure[]{
			\includegraphics[width=4.1cm]{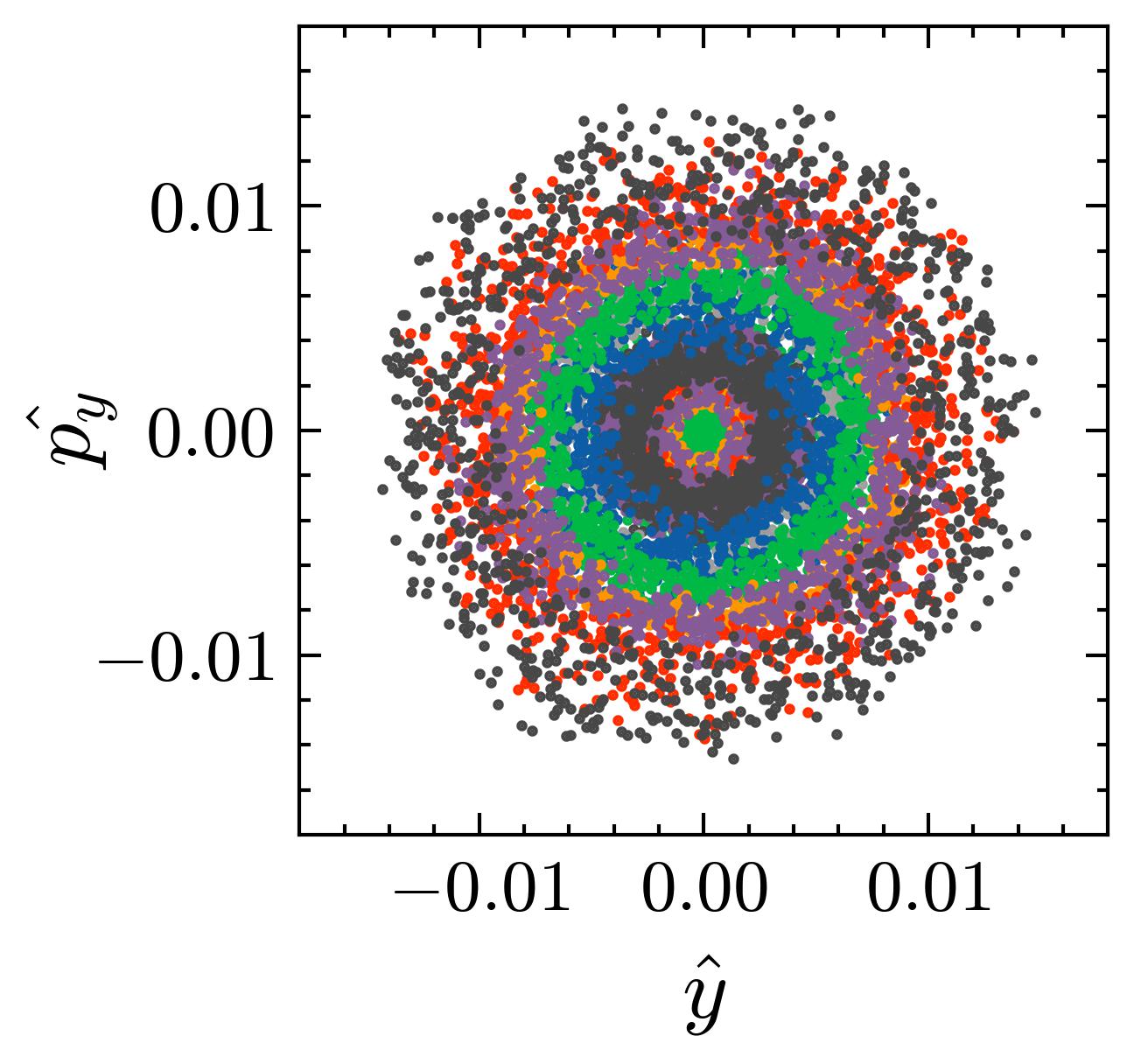}
		}
		\\
		\caption{
			\label{fig_3rd_phase_space} 
			The particle track in normalized phase spaces under the tune spread in Fig.~\ref{fig_3rd_tune_spread_coasting}. The subgraphs (a) and (b) show the the particle track in normalized phase space without errors. The subgraphs (c) and (d) show that without compensation. The subgraphs (e) and (f) show that with modified RDTs. The subgraphs (g) and (h) show that with regular RDTs. 
		}
	\end{figure}
    
	Figure.~\ref{fig_9.47_9.43_emit_chrom_off} shows the emittance evolution for the same 4 situations as previously described in Fig.~\ref{fig_half_integer_emittance}. The systematic resonance, $-4Q_x+8Q_y=36$, has a small influence on the beam, leading to a weak action transfer, about 0.7\% over 1000 turns, from the horizontal to the vertical plane even in the absence of magnetic field errors. As a result, the vertical amplitude of some particles would exceed the designed aperture even without errors, leading to 3.5\% loss of the beam in a simulation of full acceleration when the aperture is set to the designed (200,100) $\pi$ mm mrad. The compensation scheme with modified RDTs demonstrates a better performance than that with regular RDTs. With  modified RDTs, the $\varepsilon _{rms,x}$ keeps the almost same as that without errors, and the $\varepsilon _{rms,y}$ grows by 2\%, due to the insufficient compensation for halo particles at $3Q_y=28$. With  regular RDTs, the compensation does not have a good performance against these 3rd-order incoherent error driven resonances although the emittance growth is partially reduced. According to the chosen incoherent tune shift $\Delta\nu_{sc,u}<-0.3$, the difference of beta function between modified RDTs and regular RDTs has $(\beta_{sc,u}-\beta_u)/\beta_u<0.03$. The primary difference between modified RDTs and regular RDTs is still the phase. The difference becomes more significant when the on-resonance incoherent tune shift is larger. As a result, the modified RDTs compensation scheme effectively suppresses emittance growth induced by space-charge-induced incoherent error driven resonances, whereas the regular RDTs scheme does not.

	Figure.~\ref{fig_3rd_phase_space} presents the particle track in normalized phase spaces. The initial 6D coordinates of particles with the identical IDs are consistent across all four simulations. In the simulation without errors, the IDs of particles from small amplitude to large amplitude in the horizontal and vertical planes are respectively recorded as x-IDs and y-IDs. According to the 2 groups of IDs, the same particles are recorded in the rest of simulations, which means that the track of x-IDs particles is demonstrated in subgraphs (a) (c) (e) (g) and the track of y-IDs particles is demonstrated in subgraphs (b) (d) (f) (h). It can be observed that the amplitudes of particles are not invariant under such strong space charge effect, which is one issue of Eq.~(\ref{eq_tune_shift}) as previously discussed. The simulation without compensation demonstrates the serious unstable particle motion of non-core particles in the vertical plane. The effect of compensation with regular RDTs is insufficient, leading to amplitude growth of non-core particles. The unstable motion is effectively mitigated by the compensation with modified RDTs. The subgraph (f) with modified RDTs shows that the vertical amplitude growth of non-core particles caused by incoherent error driven resonances is successfully suppressed. In the subgraph (e), horizontal amplitudes of some particles increase compared to their average amplitudes in the subgraph (a), like the red particle at the edge, while that of some others decrease, like the black particle at the edge. It is difficult to summarize the transverse nonlinearity in the subgraph (e). But according to Fig.~\ref{fig_9.47_9.43_emit_chrom_off}, the horizontal RMS emittance evolution confirms that the modified RDTs compensation scheme effectively suppresses horizontal emittance growth. It can be concluded that the compensation scheme with modified RDTs successfully mitigate the beam response to these 5 3rd-order incoherent error driven resonances.
	
	In this subsection, the half-integer and 5 3rd-order incoherent error driven resonances are excited by errors and compensated by correctors in coasting-beam simulations. Both 2 groups of simulations demonstrate that the compensation scheme with modified RDTs could successfully mitigate the beam response to these incoherent error driven resonances. The feasibility of compensation scheme with modified RDTs against incoherent error driven resonances, which is the primary objective of this paper, has been validated.
	
	\subsection{Bunched-beam simulations\label{sec3_bunched}}
	In the last subsection, coasting-beam simulations without synchrotron motion have been performed and the feasibility of modified RDTs against incoherent error driven resonances has been validated. Another important issue for space charge accelerator physics is the periodic resonance crossing, which is caused by the modulation of the transverse space charge force through the synchrotron oscillations. All resonances would have additional synchrotron sidebands under periodic resonance crossing, which makes the physical picture much more complicated. 
	
	In this subsection, the simulations in Sec.\ref{sec3_coasting} would be reconstructed in the case of bunched beam in order to test the effect of modified RDTs against the periodic resonance crossing. Modified RDTs do not consider the longitudinal motion, so it can be predicted that modified RDTs are not perfect tools to describe the nonlinear behavior under periodic resonance crossing. Nonetheless, this subsection is carried out because the bunched-beam case is inevitable in a real high-intensity machine. The effect of compensation scheme with modified RDTs against the periodic resonance crossing when synchrotron motion is included must be figured out to test its feasibility in a real high-intensity machine.
	
	In simulations of this subsection, the Radio-Frequency(RF) cavity maintains a voltage of 65,000 V and a phase of 0 for the ideal particle. One bunched beam is initially generated with the RMS momentum spread $\delta = 0.0015$ and RMS longitudinal length $\sigma_s = 20 m$. The momentum spread and longitudinal length show little oscillation during simulations as they well match the RF. The intensity is reduced to maintains a suitable maximum incoherent tune shift since the peak longitudinal density has been increased while the number of macroparticles is still $1\times10^6$.
	
	The track of particles from small transverse amplitude to large transverse amplitude is no longer suitable to demonstrate the beam response due to the modulation of longitudinal motion, because the nonlinear behavior of particles with the same transverse amplitude could be different due to their different longitudinal amplitudes. In the bunched-beam case, frequency map analysis (FMA) is a good approach. FMA is a useful technique for studying the dynamics in systems described by nonlinear Hamiltonians. The motion of the system is characterized as regular or chaotic based on the evolution of its frequency in a defined time interval. In accelerator physics the main frequency of the individual particles is their incoherent tune. A full synchrotron period is chosen as the time interval in order to suppress the tune modulation induced by space charge and longitudinal effects. The evolution of tunes in both transvers planes is described by this diffusion coefficient \cite{RN57}:
	\begin{equation}
		d=\ln {\sqrt[]{(Q_x^{(2)}-Q_x^{(1)})^2+(Q_y^{(2)}-Q_y^{(1)})^2} }
		\label{eq_FMA}
	\end{equation}
	where the $Q_u^{(1)}$ and $Q_u^{(2)}$ are the tunes calculated at the first and second synchrotron period respectively. In this paper, FMA directly uses tunes of all particles, instead of on-momentum or off-momentum particles, to demonstrate the beam response.
	
	In this subsection, the 4 situations described in Sec.\ref{sec3_coasting} are constructed again. The 'without errors' and 'modified RDTs' are constructed for 20000 turns, about 125 synchrotron periods, in order to demonstrate the effect of the compensation with modified RDTs through their emittance evolution. The 'without compensation' and 'regular RDTs' are constructed for only 2 synchrotron periods in order to get their FMA. 
	
	\begin{figure}[htbp]
		\centering
		\includegraphics[width=8.6cm]{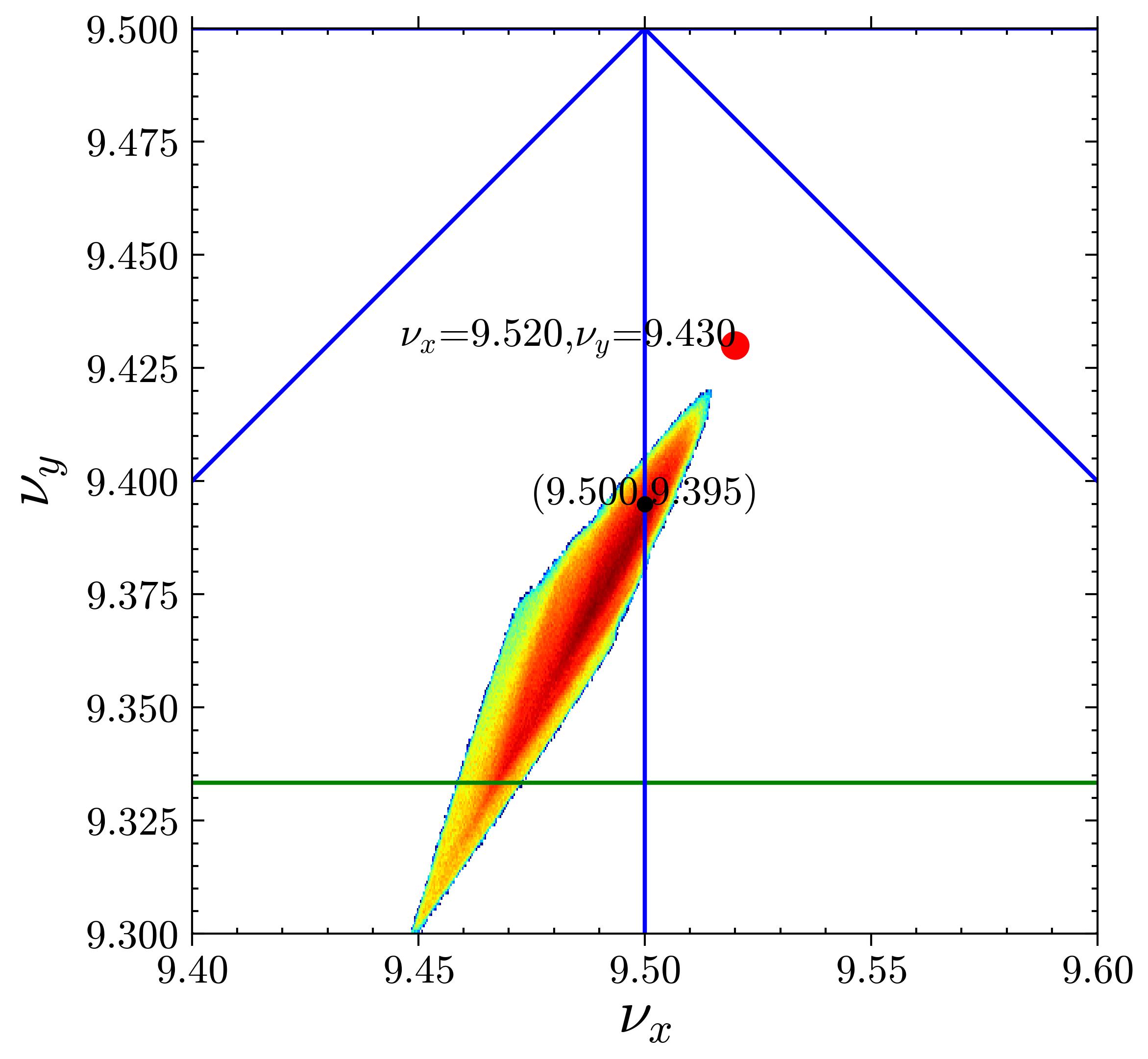}
		\caption{
			\label{fig_Half_integer_tune_spread_bunch} 
			The bunched-beam space charge tune spread map, under the resonance crossing at $2Q_{x}=19$. The black dot is one of the intersections of the tune spread and the resonance line, which could be used in Eq.~(\ref{eq_sc_phase_advance}). The tune footprint is acquired by averaging the transverse phase advance of each particle over 160 turns.
		}
	\end{figure}
	
	In the first group of simulations, the bunched-beam simulations are constructed under resonance crossing at $2Q_{x}=19$. The synchrotron tune is approximately 1/160. Fig.~\ref{fig_Half_integer_tune_spread_bunch} shows the corresponding tune spread map with reduced intensity $1.5\times10^{10}$, where the core particles are free of resonances and non-core particles would be influenced by the half-integer resonance $2Q_{x}=19$. In later simulation, the same quadrupole errors as those in Sec.\ref{sec3_coasting} on all quadrupoles, would be introduced in order to excite this half-integer resonance and corresponding compensation with modified RDTs or regular RDTs would be applied in order to test these compensation schemes against this resonance.

    \begin{figure}[htbp]
		\centering
		\includegraphics[width=8.6cm]{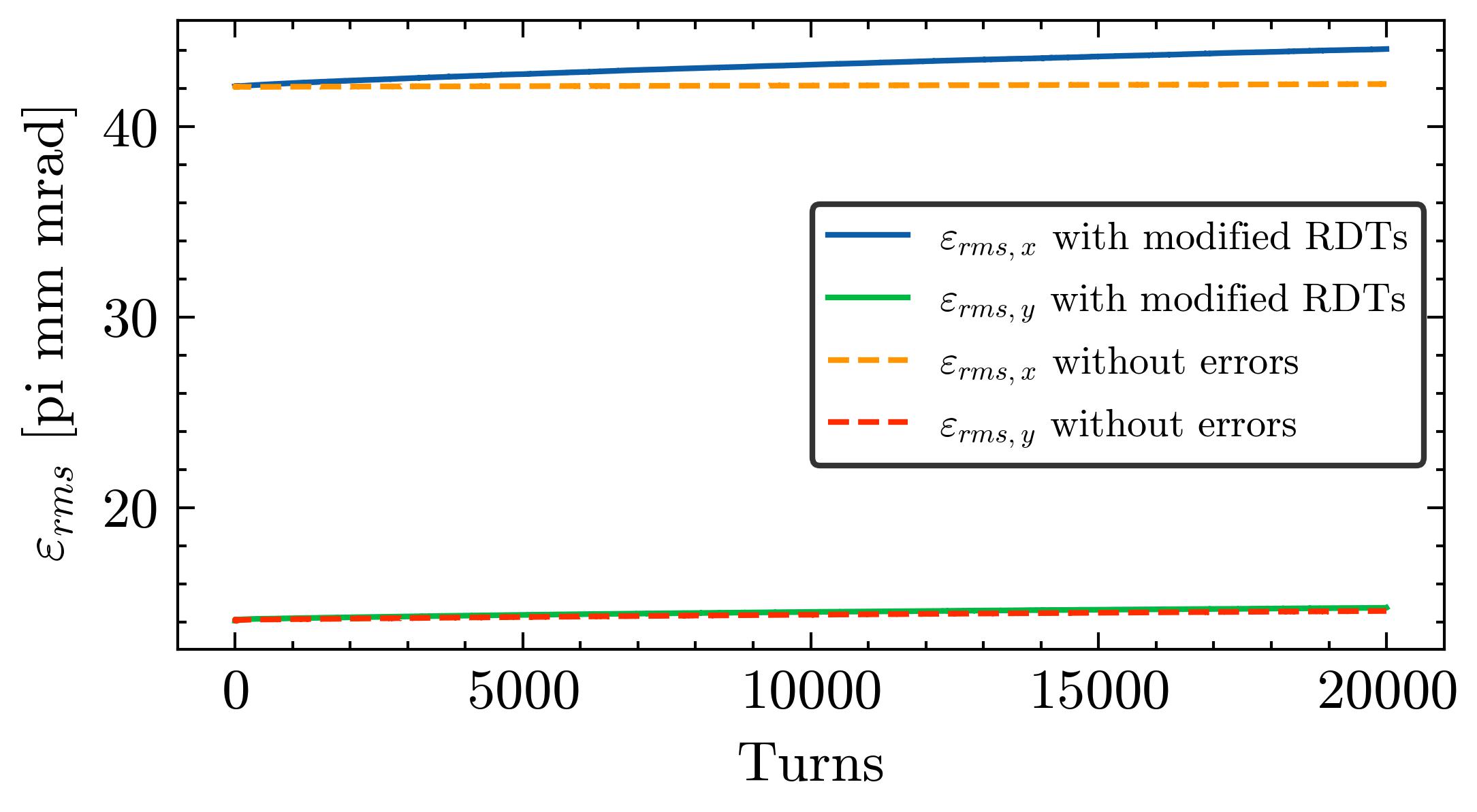}
		\caption{
			\label{fig_9.52_9.43_emit_bunch} 
			The emittance evolution under the tune spread in Fig.~\ref{fig_Half_integer_tune_spread_bunch}.
		}
	\end{figure}
    
	Figure.~\ref{fig_9.52_9.43_emit_bunch} shows the 20000-turns emittance evolution without errors and with modified RDTs. The effect of compensation scheme with modified RDTs against the half-integer resonance in bunched-beam case is not as good as that in coasting-beam case demonstrated at Sec.\ref{sec3_coasting}. The transverse emittance under compensation with modified RDTs increases by 4.3\% after 20000-turns evolution. As was predicted, compensation scheme with modified RDTs can not perfectly eliminate the influence of periodical resonance crossing. But the rate of emittance growth becomes indeed low after the compensation, which means that compensation scheme with modified RDTs against the periodical resonance crossing would still be effective in the short-term dynamical behaviors, such as capture and acceleration.
	
	\begin{figure}[htbp]
		\centering  
		\subfigure[]{
			\includegraphics[width=4.1cm]{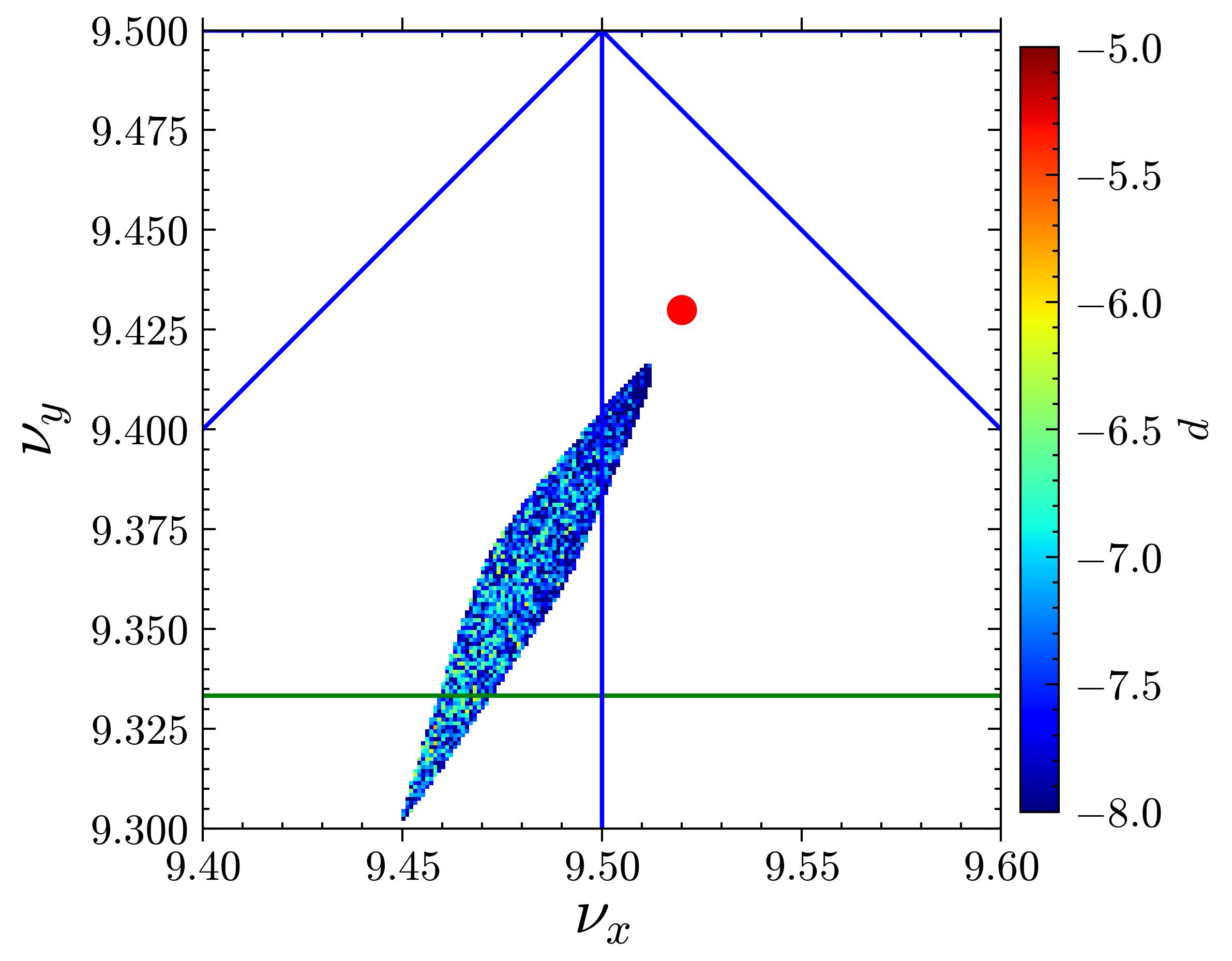}
		}
		\subfigure[]{
			\includegraphics[width=4.1cm]{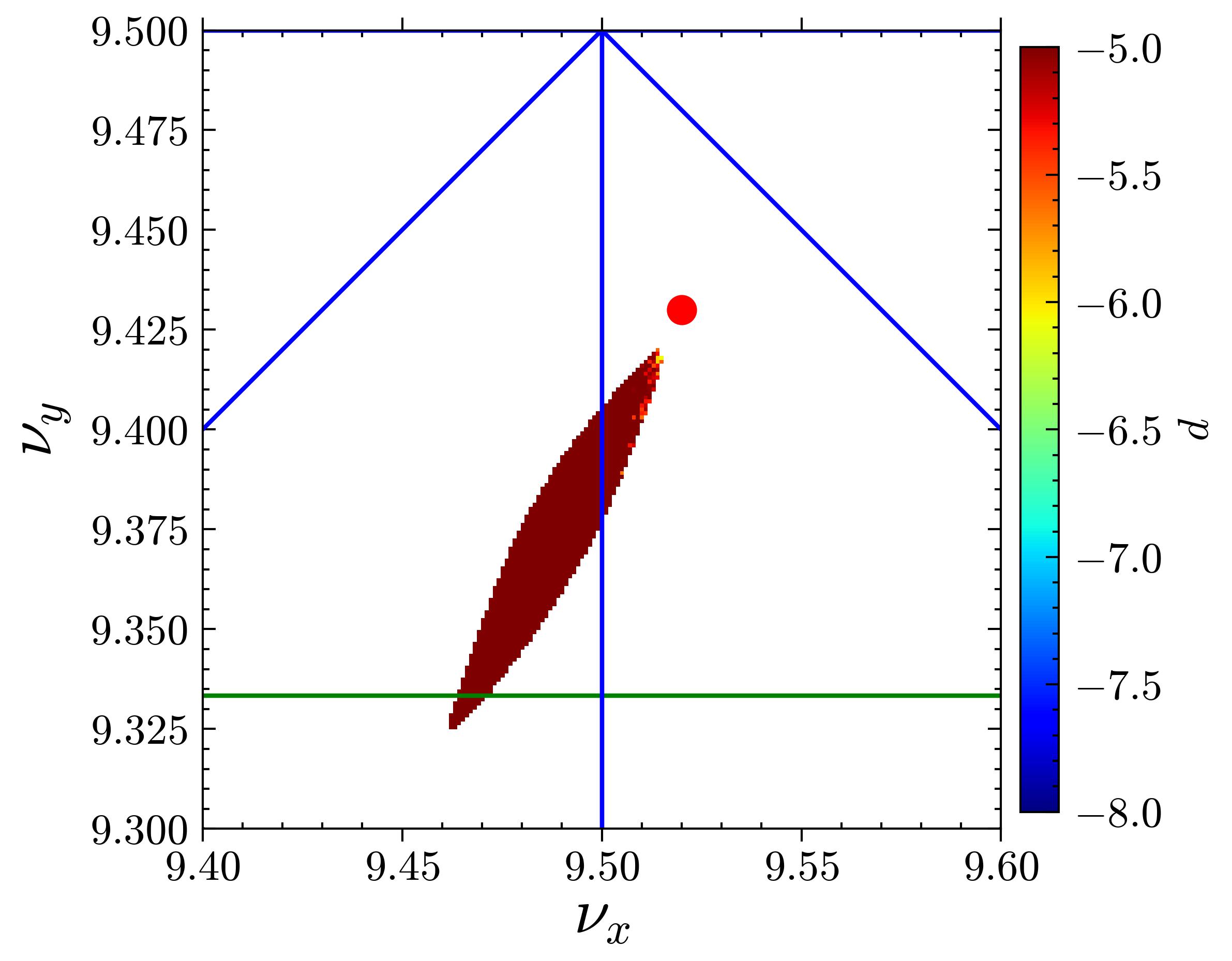}
		}
		\\
		\subfigure[]{
			\includegraphics[width=4.1cm]{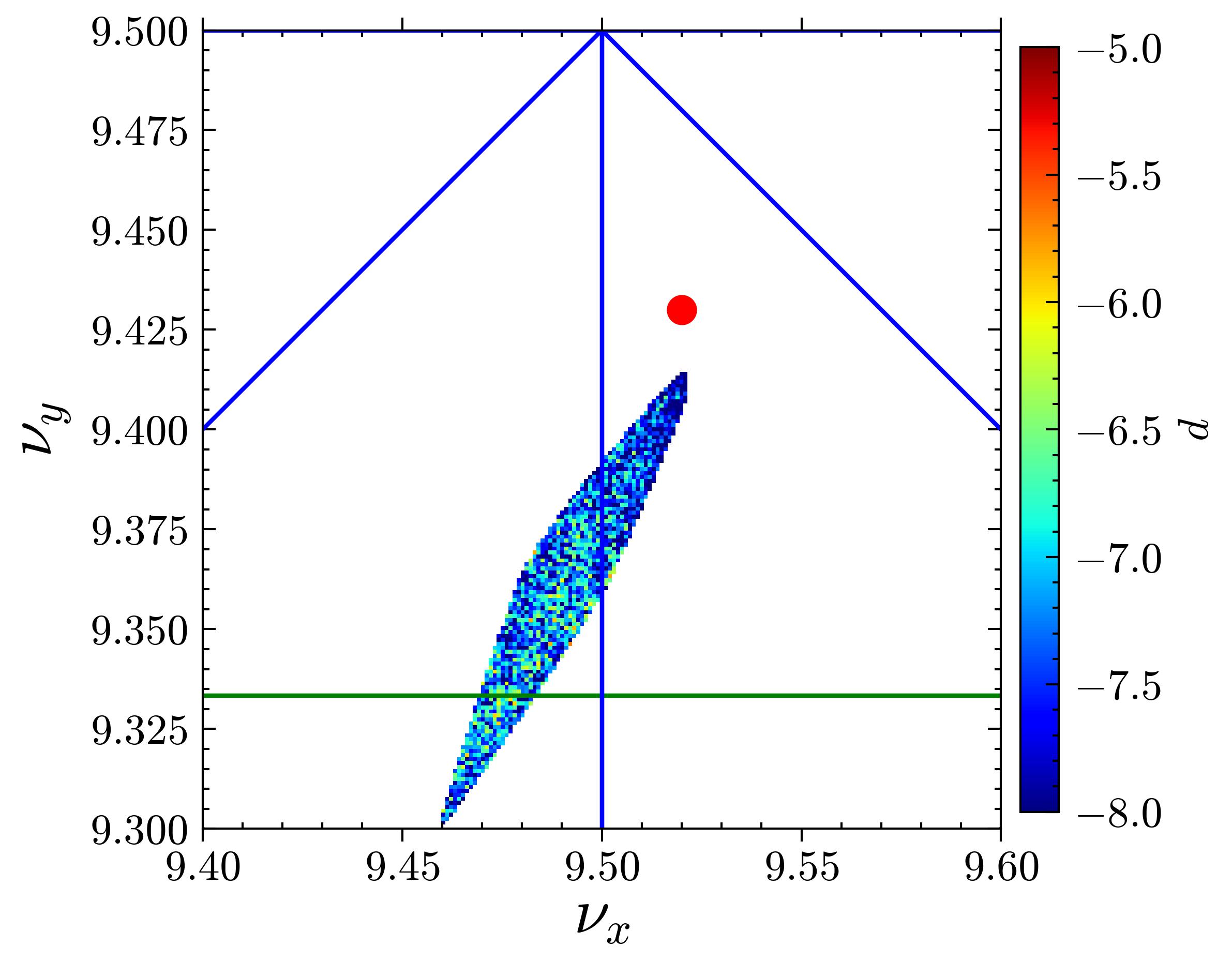}
		}
		\subfigure[]{
			\includegraphics[width=4.1cm]{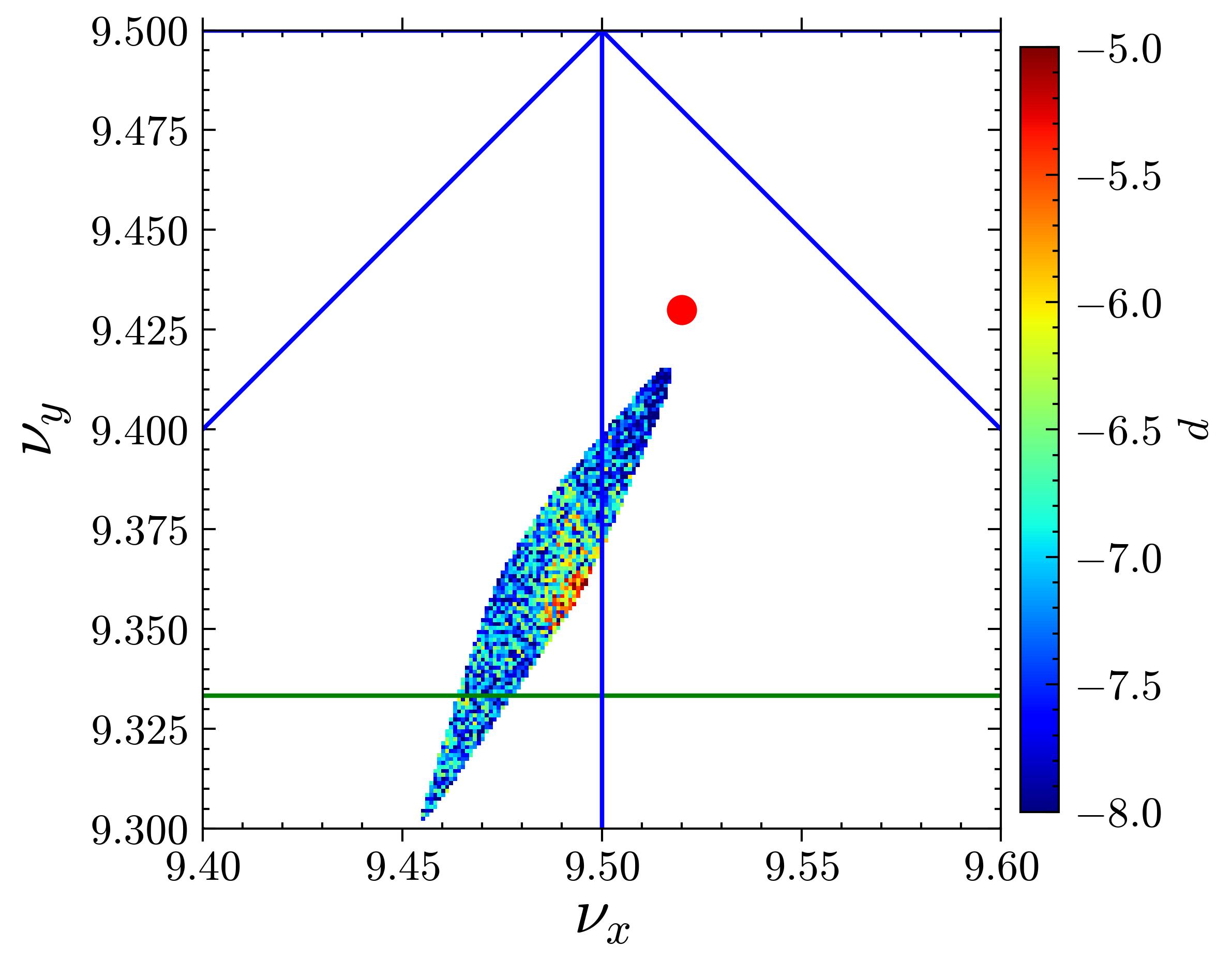}
		}
		\caption{
			\label{fig_half_FMA} 
			The FMA under the tune spread in Fig.~\ref{fig_Half_integer_tune_spread_bunch}. $Q_u^{(1)}$ of residual particles are chosen as tune coordinates of FMA. The subgraph (a) shows the FMA without errors. The subgraph (b) shows the FMA without compensation. The subgraph (c) shows the FMA with modified RDTs. The subgraph (d) shows the FMA with regular RDTs. 
		}
	\end{figure}
	
	Figure.~\ref{fig_half_FMA} shows the FMA under the tune spread in Fig.~\ref{fig_Half_integer_tune_spread_bunch}. The number of miroparticles in FMA is $1\times10^6$. In the subgraph (b), the tune spread is smaller than others because many particles have been lost during the first synchrotron period due to the nonlinearity of the half-integer resonance. And particles have been lost during the second synchrotron period. Space charge in the subgraph (b) is not constant due to beam loss. As a result, beam loss is the main reason why incoherent tunes of all particles in the subgraph (b) are modulated. In the subgraph (c), it can be observed that the bare tune has been moved a little away from the ideal bare tune (the red point) due to the combined effect of quadrupole errors and correctors. But the current tune spread still satisfies the condition where the incoherent resonance could be excited, and the beam response to this half-integer resonance is indeed suppressed by the compensation scheme with modified RDTs. In the subgraph (d), it can be observed that tunes of some on-resonance particles keep unchanged due to the compensation with regular RDTs, while tunes of some others are modulated by the half-integer resonance. By the way, the transverse emittance with regular RDTs increases by 1.7\% after 340-turns evolution. It can be concluded that the compensation scheme with modified RDTs has a good short-term performance on mitigating the beam response to the half-integer incoherent error driven resonance when synchrotron motion is included, although synchrotron motion is not considered in modified RDTs. 
	
	\begin{figure}[htbp]
		\centering
		\includegraphics[width=8.6cm]{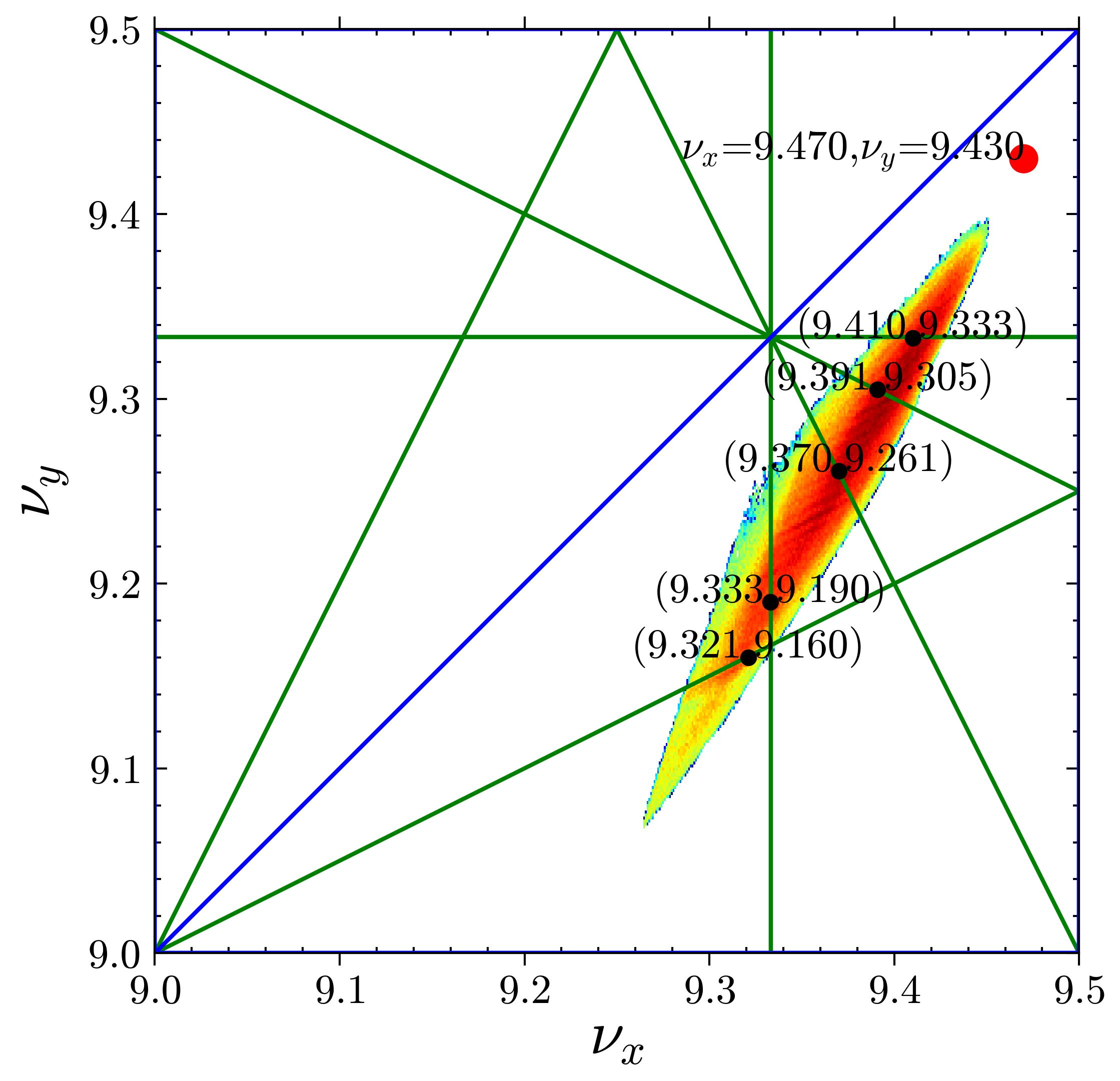}
		\caption{
			\label{fig_3rd_tune_spread_bunch} 
			The bunched-beam space charge tune spread map, under the resonance crossing at $3Q_x=28,Q_x+2Q_y=28,Q_x-2Q_y=-9$ driven by sextupole errors and $3Q_y=28,2Q_x+Q_y=28$ driven by skew sextupole errors. The black dots are intersections of the tune spread and resonance lines, which could be used in Eq.~(\ref{eq_sc_phase_advance}). The tune footprint is acquired by averaging the transverse phase advance of each particle over 157 turns.
		}
	\end{figure}
	
	In the second group of simulations, the bunched-beam simulations are constructed under resonance crossing at $3Q_x=28$, $Q_x+2Q_y=28$, $Q_x-2Q_y=-9$ driven by normal sextupoles and $3Q_y=28$, $2Q_x+Q_y=28$ driven by skew sextupoles. The synchrotron tune is approximately 1/157. Figure.~\ref{fig_3rd_tune_spread_bunch} shows the corresponding tune spread map with the intensity $4.5\times10^{10}$, where the core particles hit $3Q_x=28$, $Q_x-2Q_y=-9$ and non-core particles would be influenced by $3Q_y=28$, $2Q_x+Q_y=28$ and $Q_x+2Q_y=28$. In later simulation, the same normal and skew sextupole errors as those in Sec.\ref{sec3_coasting} on all quadrupoles, would be introduced in order to excite these resonances and corresponding compensation with modified RDTs or regular RDTs would be applied in order to test these compensation schemes against these 3rd-order resonances.
	
	\begin{figure}[htbp]
		\centering
		\includegraphics[width=8.6cm]{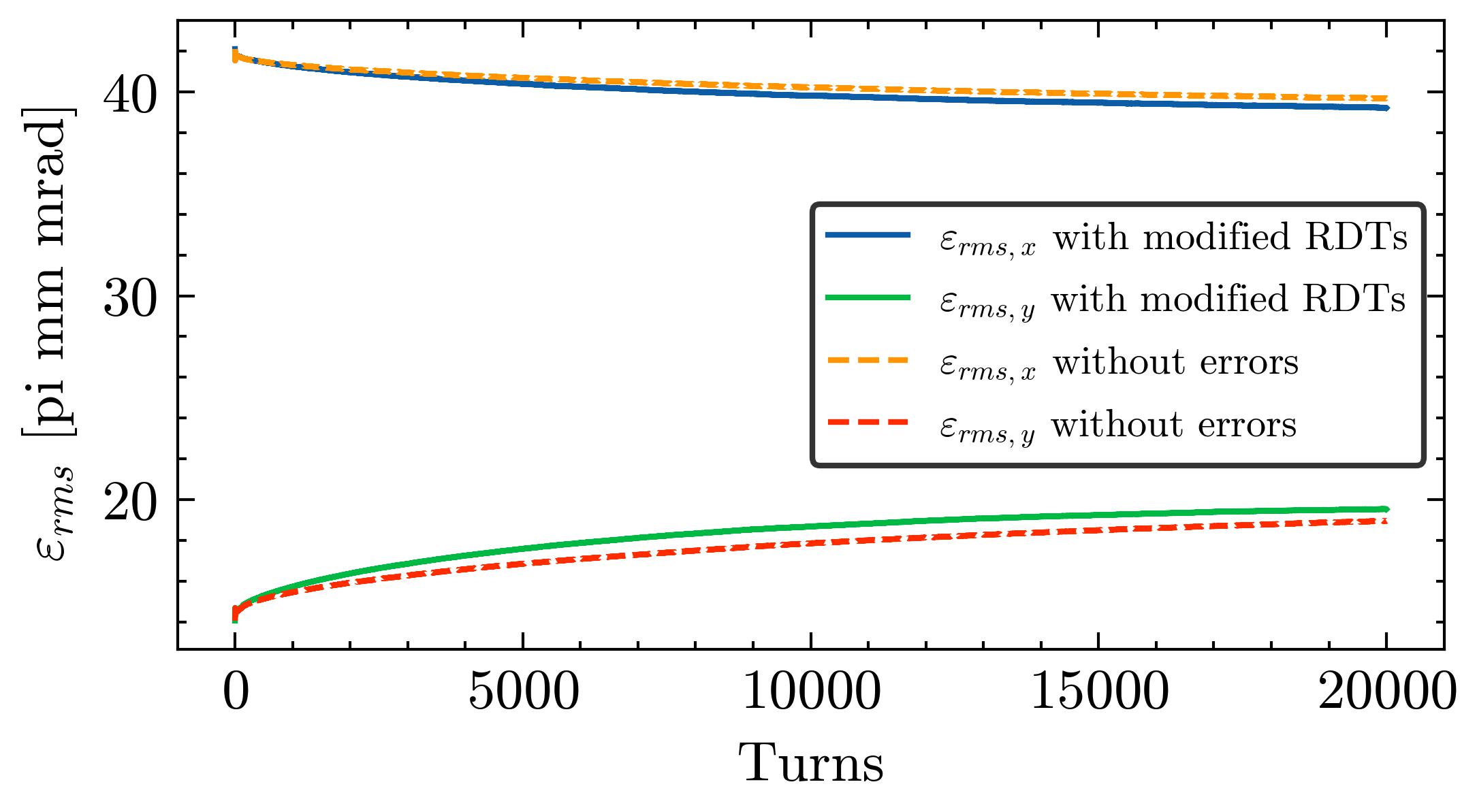}
		\caption{
			\label{fig_9.47_9.43_emit_bunch} 
			The emittance evolution whose initial tune spread is demonstrated in Fig.~\ref{fig_3rd_tune_spread_bunch}.
		}
	\end{figure}
	
	Figure.~\ref{fig_9.47_9.43_emit_bunch} shows the 20000-turns emittance evolution of 'without errors' and 'modified RDTs'. It can be observed that the emittance changes even without errors. 2 possible reasons for this phenomena are the influence of systematic resonance $-4Q_x+8Q_y=36$ and the redistribution of the Gaussian beam under such strong space charge effect. In both cases 'without errors' and 'with modified RDTs', the rate of emittance change is decreasing as the evolution. The beam is spontaneously evolving into a stable equilibrium, which means that the emittance change due to beam redistribution can be solved by optimizing the injection. The injection under strong space charge is another important issue although not discussed in this paper. The effect of compensation with modified RDTs are not perfect as was expected, leading to the additional horizontal emittance decrease of 1.3\% and additional vertical emittance increase of 3.2\%, compared to the emittance without errors at the 20000-th turn. In a high-intensity storage ring which requires the beam to evolute for a long time, if it encounters the periodical resonance crossing, this compensation scheme with modified RDTs is imperfect. 

    \begin{figure}[htbp]
		\centering  
		\subfigure[]{
			\includegraphics[width=4.1cm]{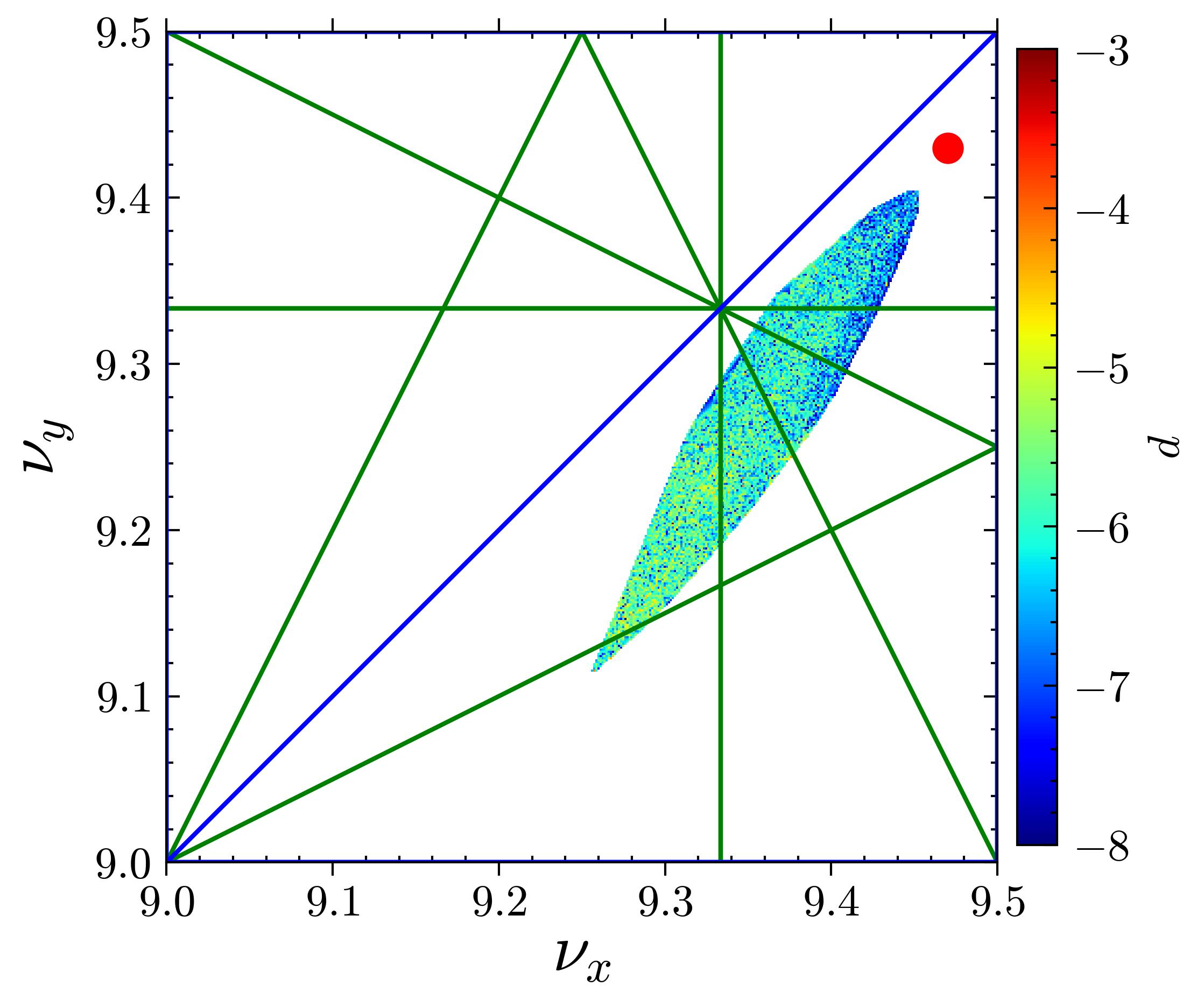}
		}
		\subfigure[]{
			\includegraphics[width=4.1cm]{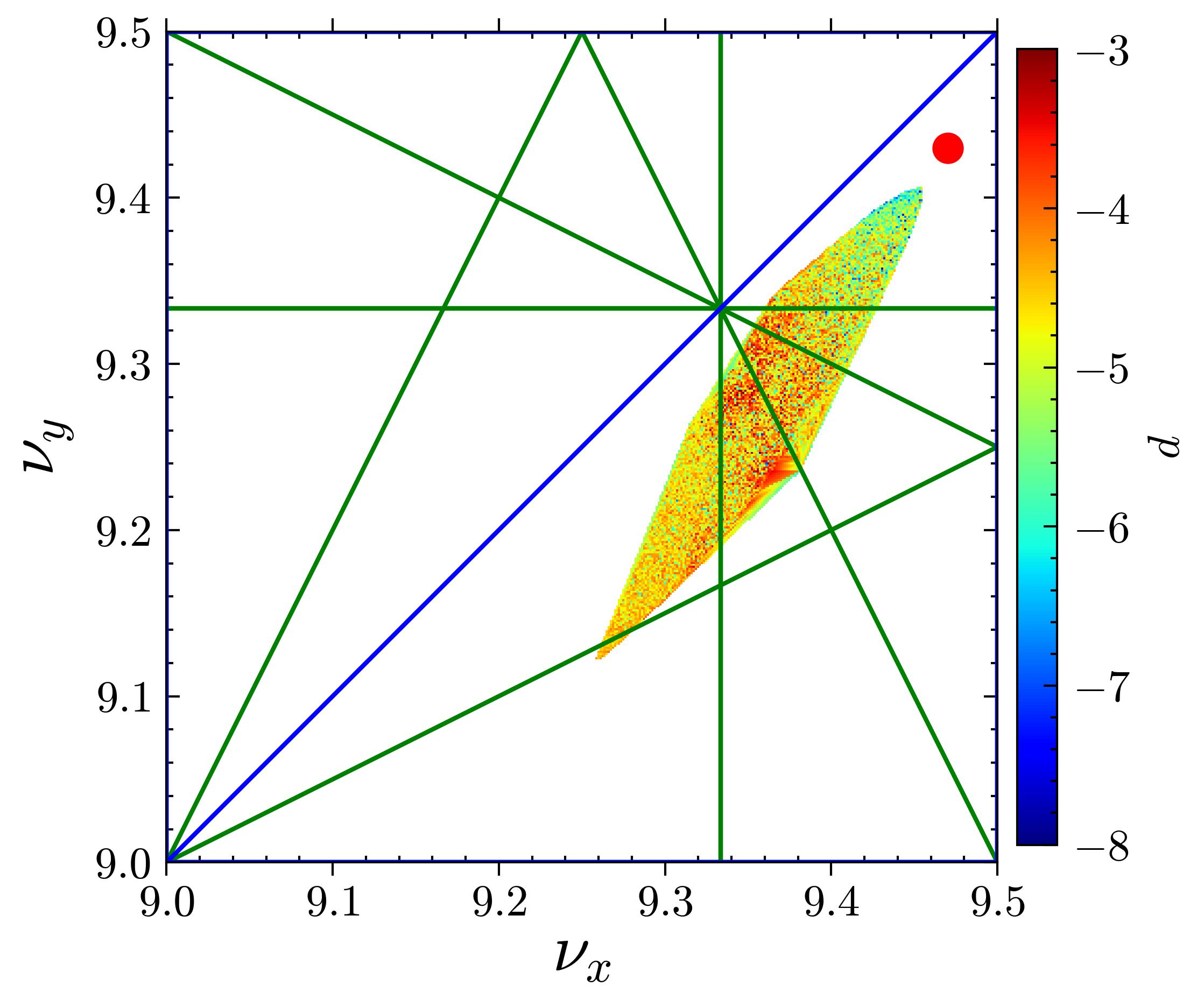}
		}
		\\
		\subfigure[]{
			\includegraphics[width=4.1cm]{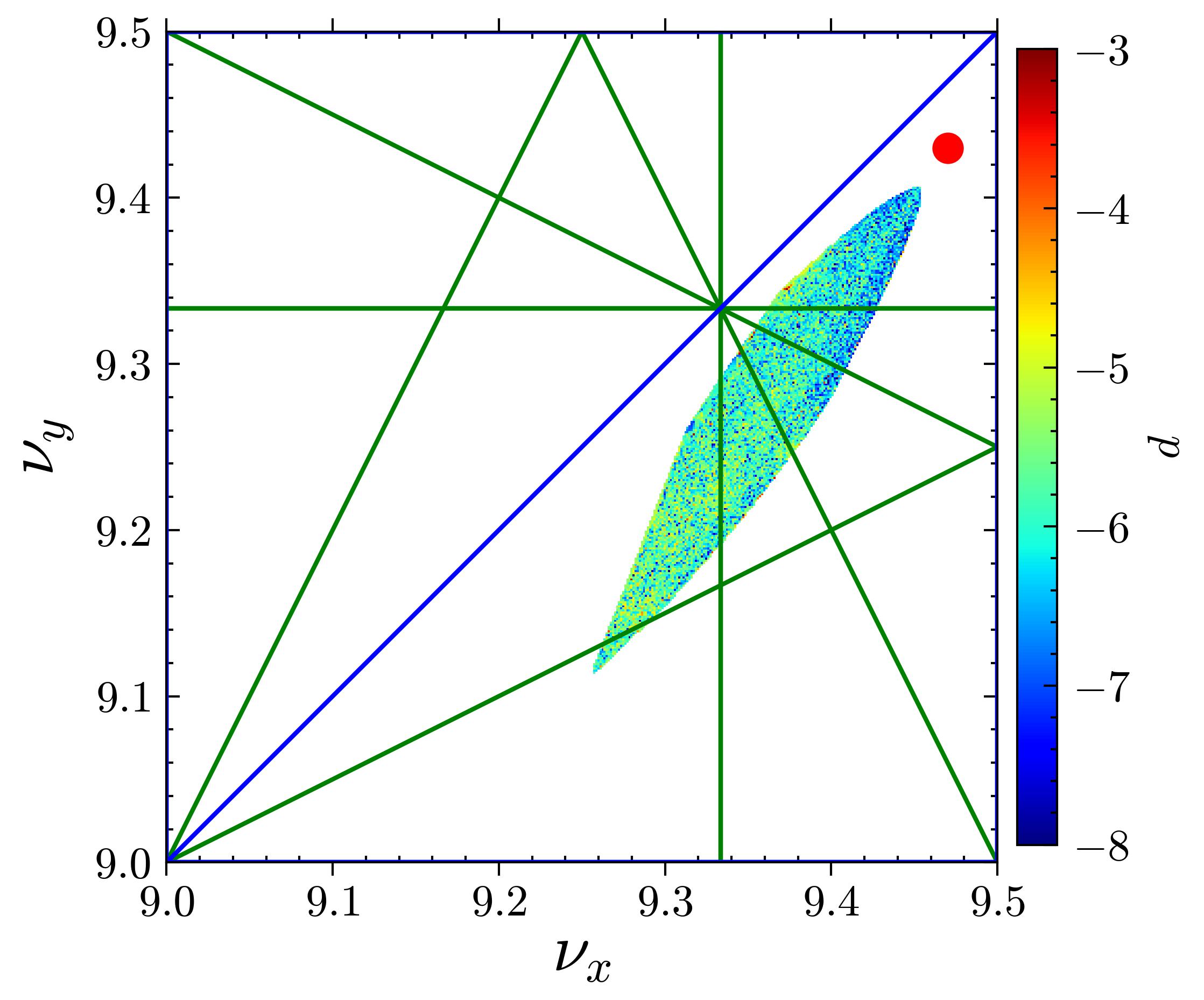}
		}
		\subfigure[]{
			\includegraphics[width=4.1cm]{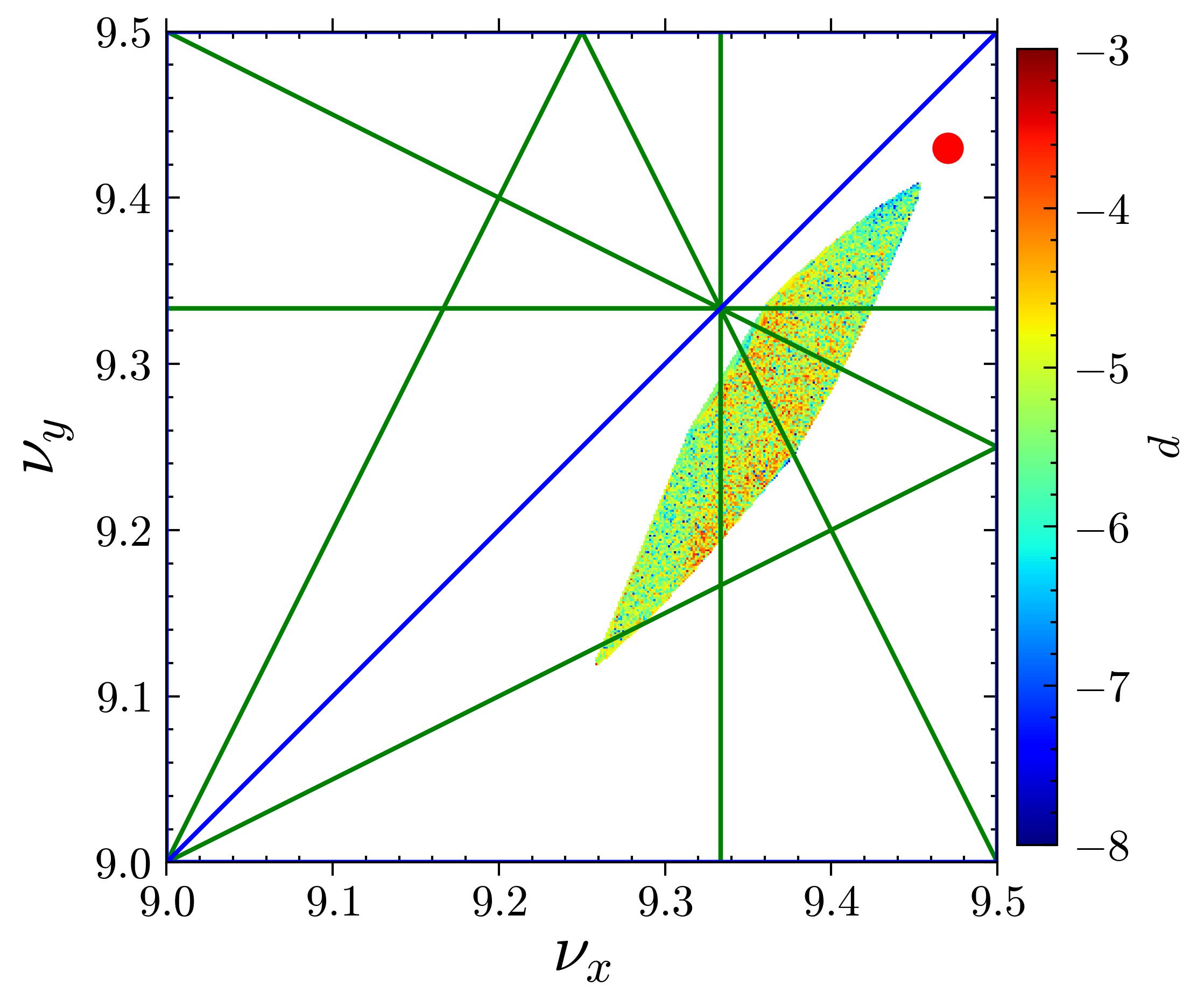}
		}
		\caption{
			\label{fig_3rd_FMA} 
			The FMA whose initial 6D coordinates are determined by the coordinates at 20000-th turn of 'without errors' in Fig.~\ref{fig_9.47_9.43_emit_bunch}. The subgraph (a) shows the FMA without errors. The subgraph (b) shows the FMA without compensation. The subgraph (c) shows the FMA with modified RDTs. The subgraph (d) shows the FMA with regular RDTs. 
		}
	\end{figure}

    Figure.~\ref{fig_3rd_FMA} shows the FMA whose initial 6D coordinates are determined by the coordinates at 20000-th turn of 'without errors' in Fig.~\ref{fig_9.47_9.43_emit_bunch}. As a result, the tune spread differs from that in Fig.~\ref{fig_3rd_tune_spread_bunch}, the transverse beam distribution is not standard Gaussian, and the redistribution would have little influence on the FMA. If the initial beam distribution in Fig.~\ref{fig_3rd_tune_spread_bunch} is directly used, the redistribution would modulate tunes of all particles even without errors, leading to confusing FMA. In the subgraph (b), the tunes of many particles are seriously modulated due to the combined effect of sextupole errors and resonance crossing. This modulation is suppressed after the compensation with modified RDTs according to the subgraph (c). The compensation with regular RDTs does not have a good performance according to the subgraph (d). Although current FMAs in Fig.~\ref{fig_half_FMA} and Fig.~\ref{fig_3rd_FMA} are noisy due to non-constant space charge model and non-single longitudinal action, they are good enough to demonstrate the suppressive effect of modified RDTs’ compensation scheme on these resonances. Besides, this simulation demonstrates that modified RDTs could be useful in a Gaussian-like beam with the elliptical symmetry. Although the compensation scheme with modified RDTs is imperfect in long-term bunched-beam case, it is good enough in short-term bunched beam case.
    
	In HIAF-BRing, the capture lasts for about 5000 turns and the vertical amplitude growth of particles would stop after 3000-turns acceleration due to the energy increase and geometric emittance decrease. In the simulation of full acceleration, the 40\% loss of the beam caused by these errors would be reduced to 3.7\% after the compensation with modified RDTs, which is almost the same as the ideal case, 3.5\%. From the view of preventing beam loss, the effect of compensation scheme with modified RDTs is good enough.
	
	In conclusion, when synchrotron motion is included, the effect of the compensation scheme with modified RDTs is good enough to prevent short-term nonlinear resonances induced by high-order magnetic fields, such as the emittance growth caused by these half-integer and 3rd-order resonances in the capture and acceleration, but it is imperfect to prevent the emittance change in a long-time storage as modified RDTs are unable to describe the nonlinear behavior under periodical resonance crossing.
	
	\section{\label{section4}CONCLUSIONS}
	In this paper, the key idea is to propose the space-charge-Twiss modification in RDTs, called modified RDTs. The modification is an approach to obtain the space-charge-induced Twiss of particles whose incoherent tunes hit resonance lines. Modified RDTs aim to compensate for non-core particles influenced by incoherent error driven resonances due to the combined effect of space charge and magnetic field imperfections. For a beam with transverse Gaussian distribution, this approach is perfect in [0, 2] $\sigma$, good but imperfect in (2, 5.6] $\sigma$, and not good in (5.6, 6] $\sigma$. The most important information of this approach is that the detuning of particles under space charge should be included in RDTs. The key difference between the space-charge-induced Twiss and the regular Twiss is the phase, which makes regular RDTs ineffective. Through minimizing modified RDTs, the compensation scheme has a good performance on mitigating the beam response to space-charge-induced incoherent error driven resonances in the coasting-beam case and corresponding periodical resonance crossing in the bunched-beam case, although modified RDTs are not a perfect dynamical explanation for periodical resonance crossing. 3 further improvement approaches, considering the longitudinal motion, the space charge potential analysis for halo particles, and the nonlinearity of space charge itself, would be introduced in further study.
	
	The feasibility of suppressing the half-integer and 3rd-order incoherent error driven resonances through modified RDTs, which is the key target of this paper, has been confirmed by coasting-beam simulations, in which the compensation scheme through minimizing regular RDTs is also demonstrated for comparison. Modified RDTs exhibits a better performance. Besides, the effect of compensation scheme with modified RDTs against the periodical resonance crossing has been tested in bunched-beam simulations. In the view of preventing beam loss and breaking through the limitations of resonance lines during the capture and acceleration, the compensation scheme is good enough. And as was expected, modified RDTs have a good but imperfect performance on mitigating the beam response to periodical resonance crossing. 
	
	Modified RDTs would be useful for high-intensity accelerators, as the compensation scheme with modified RDTs has a good performance against space-charge-induced incoherent error driven resonances. According to modified RDTs and corresponding resonance lines, magnetic correctors could be strategically placed at the appropriate phases in the lattice. By minimizing modified RDTs with these correctors, the beam loss imposed by these resonances can be mitigated. For HIAF-BRing, 40\% loss of the beam, caused by the combined effect of space-charge-induced incoherent tune spread and magnetic field imperfections, could be reduced to 3.7\% after applying the compensation scheme with modified RDTs.
	
	\section{\label{section5}Acknowledgments}
	This work is supported by the China National Funds for Distinguished Young Scientists (Grant No. 12425501).
	

\end{document}